\documentclass[twocolumn]{aastex631}
\usepackage{graphicx}
\usepackage{bm}
\usepackage{physics}

\shorttitle{BSI in AGN disk}
\shortauthors{Wang et al.}
\graphicspath{{./}{figure/}}
\newcommand{\R}[1]{\textcolor{red}{#1}}
\newcommand{\B}[1]{\textcolor{blue}{#1}}

\begin{document}
\title{Simulation of Binary-Single Interactions in AGN Disk I: Gas-Enhanced Binary Orbital Hardening} 
\author[0000-0001-5019-4729]{Mengye Wang}
\affiliation{Department of Astronomy, School of Physics, Huazhong University of Science and Technology, Luoyu Road 1037, Wuhan, China}

\author[0000-0001-7192-4874]{Yiqiu Ma}
\altaffiliation{myqphy@hust.edu.cn}
\affiliation {National Gravitation Laboratory, School of Physics, Huazhong University of Science and Technology, Luoyu Road 1037, Wuhan, China}
\affiliation{Department of Astronomy, School of Physics, Huazhong University of Science and Technology, Luoyu Road 1037, Wuhan, China}

\author[0000-0003-3556-6568]{Hui Li}
\affiliation{Theoretical Division, Los Alamos National Laboratory, Los Alamos, NM 87545, USA}

\author[0000-0003-4773-4987]{Qingwen Wu}
\altaffiliation{qwwu@hust.edu.cn} 
\affiliation{Department of Astronomy, School of Physics, Huazhong University of Science and Technology, Luoyu Road 1037, Wuhan, China}

\author[0000-0002-7329-9344]{Ya-Ping Li}
\affiliation{Shanghai Astronomical Observatory, Chinese Academy of Sciences, Shanghai 200030, People’s Republic of China}

\author[0009-0003-0516-5074]{Xiangli Lei}
\affiliation{Department of Astronomy, School of Physics, Huazhong University of Science and Technology, Luoyu Road 1037, Wuhan, China}

\author[0000-0002-2581-8154]{Jiancheng Wu}
\affiliation{Department of Astronomy, School of Physics, Huazhong University of Science and Technology, Luoyu Road 1037, Wuhan, China}

\begin{abstract}
Stellar-mass binary black hole\,(BBH) mergers within the accretion disks of active galactic nuclei may contribute to gravitational wave\,(GW) events detected by grounded-based GW detectors. In particular, the interaction between a BBH and a single stellar-mass black hole\,(sBH), known as the binary-single interaction\,(BSI) process,  can potentially lead to GW events with detectable non-zero eccentricity. Previous studies of the BSI process, which neglected the effects of gas,  showed that BSIs contribute non-negligibly to GW events in a coplanar disk environment. In this work, we conduct a series of 2-dimensional hydrodynamical and N-body simulations to explore the BSI in a gas environment by coupling REBOUND with Athena++. We perform 360 simulation runs, spanning parameters in disk surface density \(\Sigma_0\) and impact parameter \(b\). We find that the gas-induced energy dissipation within the three-body system becomes significant if the encounter velocity between the sBHs is sufficiently large\,($\gg c_s$).  Our simulation results indicate that approximately half of the end states of the BSI are changed by gas. Furthermore, at higher gas density, the number of close encounters during the BSI process will increase and the end-state BBHs tend to be more compact. Consequently, the presence of gas may shorten the GW merger timescale for end-state BBHs and increase the three-body merger rate.

\end{abstract}


\section{Introduction}    \label{sec:intro}
Up to now, more than 90 binary black hole\,(BBH) mergers have been detected by ground-based gravitational wave\,(GW) detectors such as LIGO, Virgo, and KAGRA\,\citep[][]{Acernese2015,LIGO2015,Kagra2019NatAs}. More GW merger events will be detected as the O4/O5 detection runs complete. The formation channels of these merger events are still a subject that requires further research. Two main BBH formation channels are the isolated evolution of massive binaries in galactic fields\,\citep[e.g.,][]{Bethe1998ApJ,Belczynski2016Natur,Stevenson2017NatCo,Giacobbo2018MNRAS} and the dynamical evolution in dense stellar environments, such as  globular clusters\,\citep[][]{Zwart2000ApJ,Ziosi2014MNRAS,Rodriguez2016ApJ,Barber2024arXiv}, nuclear star clusters\,\citep[][]{Antonini2019MNRAS,Kritos2022PhRvD}, and accretion discs around supermassive black holes\,(SMBHs) in active galactic nuclei\,\citep[AGN, e.g.,][]{McKernan2014MNRAS,Bartos2017NatCo,stone2017MNRAS,Yang2019PhRvL,Tagawa2020ApJ}. So far, there are some indications that at least two GW mergers may have originated from AGN channels: GW190521\,\citep[][]{Abbott2020PhRvL,Abbott2020ApJL} and GW170817A\,\citep[][]{Zackay2021PhRvD}. GW190521, with a total mass of approximately \( 150 \, M_\odot \), could be accompanied by an electromagnetic counterpart from a known AGN observed by the ZTF\,\citep[][]{Graham2020PhRvL,Chen2022MNRAS}, though this association remains controversial. The chirp mass $\sim 40\,M_\odot$ and effective spin $\sim 0.4$ of GW170817A has also been argued as coming from a hierarchical merger in an AGN disk\,\citep[][]{Gayathri2020ApJ}.


The most distinctive feature of AGN disk, as a potential factory for BBH mergers, is the presence of a dense gas environment. Both the in-situ star formation in the outer gravitationally unstable region of the AGN disk\,\citep[][]{SG2003MNRAS,TQM2005ApJ} and the capture of stellar-mass black hole\,(sBH) by the AGN disk\,\citep[][]{Vilkoviskij2002A&A,Kennedy2016MNRAS,Wang2023MNRAS} will increase the number of sBHs in the AGN disk. These sBHs, embedded in the disk, will grow through accretion and migrate towards the central SMBH due to gas torques\,\citep[e.g.,][]{Goldreich1979ApJ,Tanaka2002ApJ,LYP2024ApJ}.  Due to the existence of migration trap, the probability of gravitational encounters of these sBHs will increase a lot\,\citep[e.g.,][]{Masset2003ApJ,Lyra2010ApJ,McKernan2012MNRAS,Bellovary2016ApJ,Secunda2019ApJ}. Recently, hydrodynamical simulation works have shown that the presence of gas in  the AGN disk will facilitate the BBH formation and accelerate their mergers\,\citep[e.g.,][]{Baruteau2011ApJ,LYP2022ApJ,LRX2022MNRAS,Dempsey2022ApJ,LJR2023ApJ,Rozner2023MNRAS,Rowan2023MNRAS,Rowan2024arXiv,Whitehead2024MNRAS,Whitehead2024MNRAS_disk_nova,Dodici2024ApJ}.

Furthermore, a BBH can gravitationally interact with a single sBH, which is the so-called Binary-Single Interaction\,(BSI) process\,\citep[see][for a review]{Valtonen2006tbp}. Previous studies have shown that these BSI process can induce eccentric BBH mergers in globular clusters\,\citep[e.g.,][]{Gultekin2006ApJ,Rodriguez2016PhRvD,Samsing2018PhRvD,Giant2021MNRAS,Parischewsky2023ScPPC,Ginat2023MNRAS} or gas-rich clusters \,\citep[][]{Rozner2022ApJ}. However,  the merger rate resulting from the BSI in the cluster is much lower than that of direct BBH mergers.
Interestingly, this scenario will change by introducing a coplanar system such as an AGN disk environment. Recently, \cite{Samsing2022Natur} suggested that the coplanar AGN disk environment may enhance the fraction of BSI-induced merger by two orders of magnitude compared to that in the spherical cluster case\,\citep[see also][]{Valtonen2006tbp}. Thus, the  BSI-induced merger rate  in AGN disk can be at least comparable to that of direct BBH mergers, which may provide a non-negligible channel for BBH mergers with non-zero eccentricity\,\citep[such as GW190521, see][]{Graham2020PhRvL} or hierarchical triple mergers\,\citep[see][]{Li-Fan2025arXiv}. Next-generation ground-based GW detectors are expected to have a high sensitivity for resolving the orbital eccentricity of BBH mergers\,\citep[e.g.,][]{Saini2024MNRAS} and have a large number of detection events, which will serve as a crucial diagnostic tool for distinguishing different formation channels of these events.


There have been several works about the BSI process in the AGN disk\,\citep[e.g.,][]{Samsing2022Natur,Trani2024A&A,Fabj2024arXiv}, which suggested that some detected BBH merger events may attribute to the BSI process in AGN disk. All these studies were conducted using analytical methods or N-body simulations, neglecting the role of gas. However, the presence of gas can interact with the three-body system during the BSI process, significantly influencing the end state of the BSI. In this work, we combine 2-dimensional\,(2D) hydrodynamical simulations of gas dynamics with N-body simulations of three-sBH interactions to investigate how dense gas within an AGN disk influences the BSI processes. By conducting 360 simulations across a range of gas densities and impact parameters, we analyze the detailed energy and angular momentum dissipation in the three-body system, demonstrating that the presence of gas effectively dissipates its energy. Furthermore, we statistically compare the orbital parameters of the end-state BBHs under varying disk densities, revealing that the semi-major axis of the end-state BBH tends to form a more compact distribution.

The structure of this paper is as follows. Section\,\ref{sec:model} presents the numerical methods, simulation tools, and initial parameter settings. In Section\,\ref{sec:Fiducial_results}, we present our simulation results for a fiducial model to understand the BSI process in a typical three-body system in an AGN disk. This section will include a description of the gas morphology around the sBHs, along with the energy and angular momentum dissipation. Section\,\ref{sec:parameter_space} is devoted to the statistical analysis of our 360 simulation runs. Finally, we summarize our work with some discussions in Section\,\ref{sec:discussions}.

 \begin{figure*}
\centering
\includegraphics[scale=0.4]{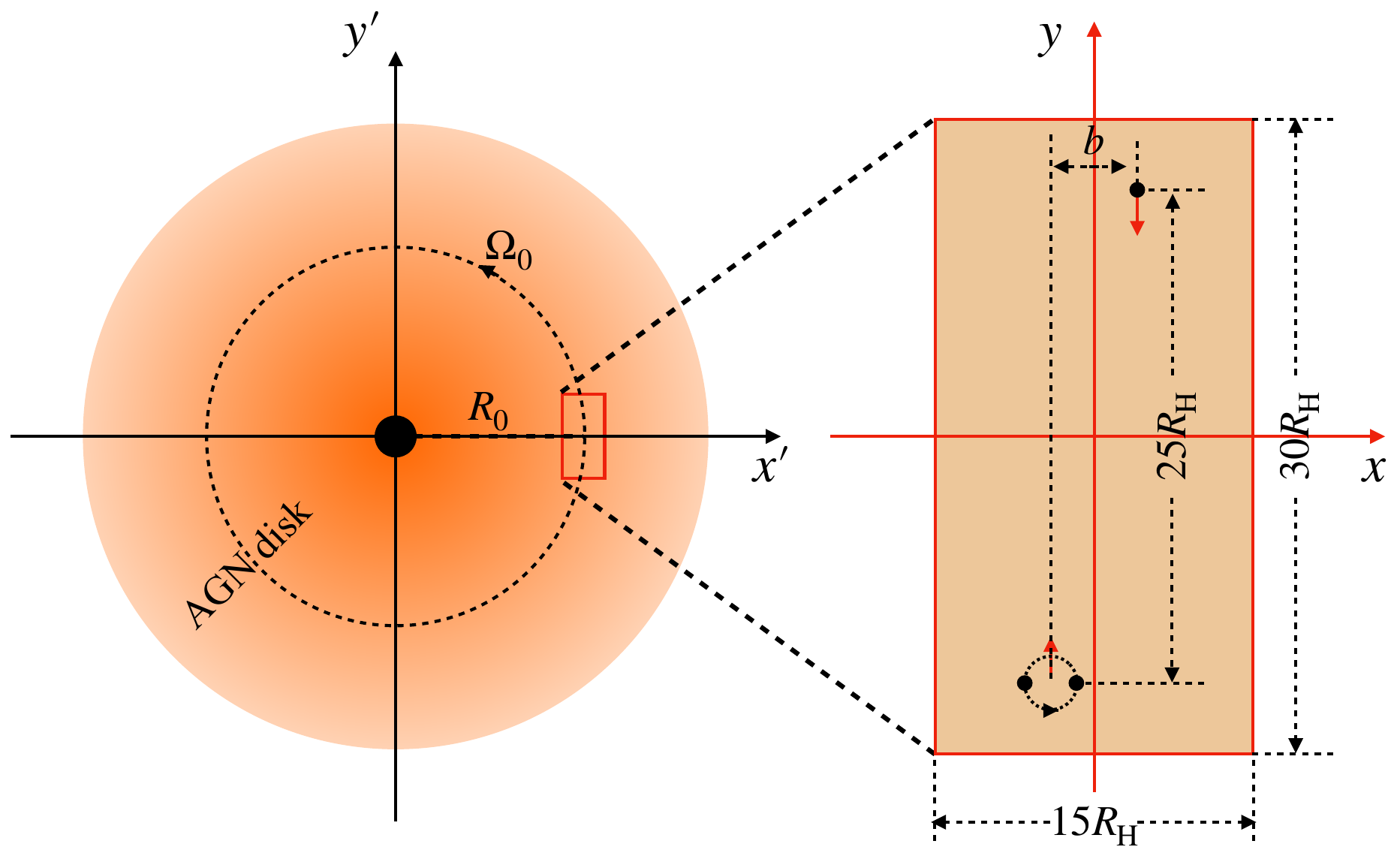}
\caption{Schematic representation of the physical scenario. We take a 2D rectangular patch\,(shearing box) from an AGN disk that is in Keplerian rotation around the SMBH. The shearing box is located at a distance $R_0$ from the SMBH and performs circular motion around the SMBH with an angular frequency $\Omega_0$. To save computational resources, we only simulate the hydrodynamics of the shearing box in the corotating frame $(x,\, y)$. A BBH and a single sBH are embedded in the disk. The initial binary is configured to be prograde. The size of the shearing box is set as $15\,R_{\rm H}\times 30\,R_{\rm H}$. The distance from the center of mass of the BBH to the single sBH in the $x$ and $y$ directions is set to $b$ and $25\,R_{\rm H}$, respectively. }
\label{fig:sch_rep}
\end{figure*}

\section{METHOD} \label{sec:model}
The background environment of our system consists of an SMBH of mass $M$ and a Keplerian gas disk surrounding the SMBH. As shown in Figure\,\ref{fig:sch_rep}, we track a 2D rectangular patch on the AGN disk\,(shearing box) at a distance $R_0$ from the SMBH. Inside the AGN disk, a BBH and a single sBH of mass $m_1=m_2=m_3$ are embedded. Both the BBH's center of mass and the single sBH are orbiting around the SMBH with circular Keplerian orbits, while the two sBH components of the BBH are circularly orbiting around their center of mass. Moreover, we set the initial BBH to be prograde, meaning the angular momentum of the BBH around its center of mass is aligned with the orbital angular momentum around the SMBH. The characteristic length scale in the shearing box is defined to be the Hill radius of the three sBHs,
\begin{equation}
    R_{\rm H} = R_0 \left(\frac{M_{\rm sBH}}{3M}\right)^{1/3},
\end{equation}
where the total mass of the three sBHs is $M_{\rm sBH}=m_1+m_2+m_3$. 
In the following, we will introduce the basic physical processes of our simulation in the corotating frame ($x,\,y$).

\subsection{Gas Dynamics}
In the shearing box, the time evolution of the gas velocity $\bm{u}$ and gas surface density, $\Sigma$ are described by the Navier-Stokes equations:
\begin{equation}
\begin{split}
  &  \frac{\partial \Sigma}{\partial t} + \nabla \cdot (\Sigma \bm{u}) = 0, \\
  & \frac{\partial (\Sigma \bm{u}) }{\partial t} + \nabla\cdot (\Sigma \bm{u}\bm{u}) + \nabla P = \Sigma(\bm{a}_{\rm SMBH} + \bm{a}_{\rm vis} + \bm{a}_{\rm BH}),
\end{split}
\end{equation}
where $\Sigma$, $\bm{u}$ and $P$ are gas surface density, velocity and pressure, respectively. In the corotating frame, the acceleration of the Keplerian motion around the SMBH $\bm{a}_{\rm SMBH}$, is given by:
\begin{equation}
    \bm{a}_{\rm SMBH} = 2\bm{u}\times (\Omega_0\hat{z}) + 2q\Omega^2_0 x\hat{x},
\end{equation}
where $q \equiv -{\rm d\,ln}\, \Omega/ {\rm d\,ln}\, R$ is the background shear parameter and is 3/2 for a Keplerian disk and $\Omega_0\equiv(GM/R_0^3)^{1/2}$ is the angular velocity of the corotating frame. The $ \hat x, \hat y$ are the unit vectors in the corotating frame as shown in Figure\,\ref{fig:sch_rep}, while $\hat z$ is the unit vector perpendicular to the $xy$ plane.
The viscous acceleration $\bm{a}_{\rm vis}$ is the divergence of viscous stress tensor
\begin{equation}
    \bm{a}_{\rm vis} = \nabla \cdot \bm{T},
\end{equation}
with
\begin{equation}
    T_{ij} = \nu \Sigma \left( \frac{\partial u_i}{\partial x_j} + \frac{\partial u_j}{\partial x_i} - \frac{2}{3}\delta_{ij}\nabla \cdot \bm{u}  \right),
\end{equation}
where $\nu$ is the kinematic viscosity and we adopt $\alpha$-viscosity so that $\nu = \alpha c_s H$ with $c_s$ and $H$ are sound speed and disk scale height.
Finally, the gas also gravitationally interacts with the embedded sBHs with acceleration given by
\begin{equation}
    \bm{a}_{\rm BH} = - \nabla \phi_{\rm BH} (\bm{r}), 
\end{equation}
and
\begin{equation}
    \phi_{\rm BH} (\bm{r}) = -\sum_{k=1}^3 \frac{Gm_{k}}{(|\bm{r}-\bm{r}_k|^2+\epsilon^2)^{1/2}},
\end{equation}
where $m_k$ and $\bm{r}_k$ are the mass and location of the $k_{\rm th}$ sBH, $\bm{r}$ specifies the location of the fluid element, and $\epsilon$ is the gravitational softening length introduced to regularize the divergence of the gravitational potential. To model the accretion of the gas into the sBH, a simple mass removal algorithm is used so that the gas will be removed when gas enters within a radius $r_{\rm sink}$ near the sBH\,\citep[e.g.,][]{LYP2022ApJ}. The gas-removal rate is  $\dot{\Sigma}/\Sigma  = n_{\rm b} \Omega_{\rm b0}$, where $n_{\rm b}$ is a free parameter and $\Omega_{\rm b0}=\sqrt{G(m_1+m_2)/a_{\rm b0}^3}$ is the initial orbital angular frequency of the BBH.
Consistent with the previous studies\,\citep[e.g.,][]{LYP2021ApJ,LYP2024ApJ}, we neglect the changes in mass and momentum of the sBHs due to gas accretion. We discuss this issue in detail in Section\,\ref{sec:discussions}.

\subsection{N-body Dynamics}
We consider the dynamic process between a BBH and a single sBH influenced by the background gravitational potential of the center SMBH and the AGN disk. When the distance between BBH and the single sBH is less than $R_{\rm H}$, they will suffer a strong scattering. The BSI interaction will boost the eccentricity of an initially circular BBH system to a higher value. In this work, to understand the role of the dense gas in the BSI process, we will simulate the gravitational dynamics of the three sBHs and the central SMBH using the N-body IAS15\,\citep{Everhart1985ASSL,Rein2015MNRAS} algorithm in REBOUND\,\citep{Rein2012A&A}. The gravitational force on the sBHs contributed by the dense gas can be written as,
\begin{equation} \label{eq:a_gas}
    \bm{a}_{k,\rm gas} = G\sum_{i=1}^{n_x} \sum_{j=1}^{n_y} {\rm d}m_{ij}\, \frac{\bm{r}_{ij}-\bm{r}_{k}}{ (|\bm{r}_{ij}-\bm{r}_{k}|^2 + \epsilon^2)^{(3/2)} },
\end{equation}
where $k$ = \{1,\,2,\,3\} denote the $k^{\rm th}$ sBH, $n_x$ and $n_y$ are the number of fluid cells in the $x$-direction and $y$-direction, ${\rm d}\,m_{ij}$ and $\bm{r}_{ij}$ are the mass and location of the $(i^{\rm th},\, j^{\rm th})$ fluid cell, $\bm{r}_k$ is the location of the $k^{\rm th}$ sBH. 

The binary orbital dynamics can be characterized by the orbital energy $E_{\rm b}$ and angular momentum $L_{\rm b}$:
\begin{equation}  \label{eq:energy_binary}
\begin{split}
    &E_{\rm b} = \frac{1}{2} \mu |\bm{v}_1-\bm{v}_2|^2 - \frac{Gm_{\rm b}\mu}{|\bm{r}_1-\bm{r}_2|},\\
    &L_{\rm b} = \mu(\bm{r}_1-\bm{r}_2)\times (\bm{v}_1-\bm{v}_2),
\end{split}
\end{equation}
where $\bm{v}_k$, $\bm{r}_k$ are the velocities and positions of the sBHs $k=(1,\,2)$, $m_{\rm b} = m_1+m_2$ and $\mu = m_1m_2/(m_1+m_2)$ are the total mass and the reduced mass of the binary system. Meanwhile, the semi-major axis and eccentricity of the binary can be evaluated via
\begin{equation}\label{eq:BBH_orbital_parameters}
    a_{\rm b} = -\frac{Gm_{\rm b}\mu}{2E_{\rm b}},\quad
    e_{\rm b} = \sqrt{ 1+\frac{2E_{\rm b}L_{\rm b}^2}{G^2m_{\rm b}^2\mu^3}}.
\end{equation}

For a three-body system, the total energy and angular momentum are
\begin{equation} \label{eq:total_EL}
\begin{split}
    &E = \frac{1}{2}\sum_{k=1}^3 m_k |\bm{v}_k- \bm{v_c}|^2 - \sum_{1\leq k<l\leq n}^3 \frac{Gm_km_l}{|\bm{r}_k-\bm{r}_l|},\\
    &L = \sum_{k=1}^3 m_k(\bm{r}_k-\bm{r}_c)\times (\bm{v}_k-\bm{v}_c),
\end{split}
\end{equation}
where $\bm{r}_c$ and $\bm{v}_c$ is the center-of-mass position and velocity of three-body system.

\begin{figure}
\centering
\includegraphics[scale=0.4]{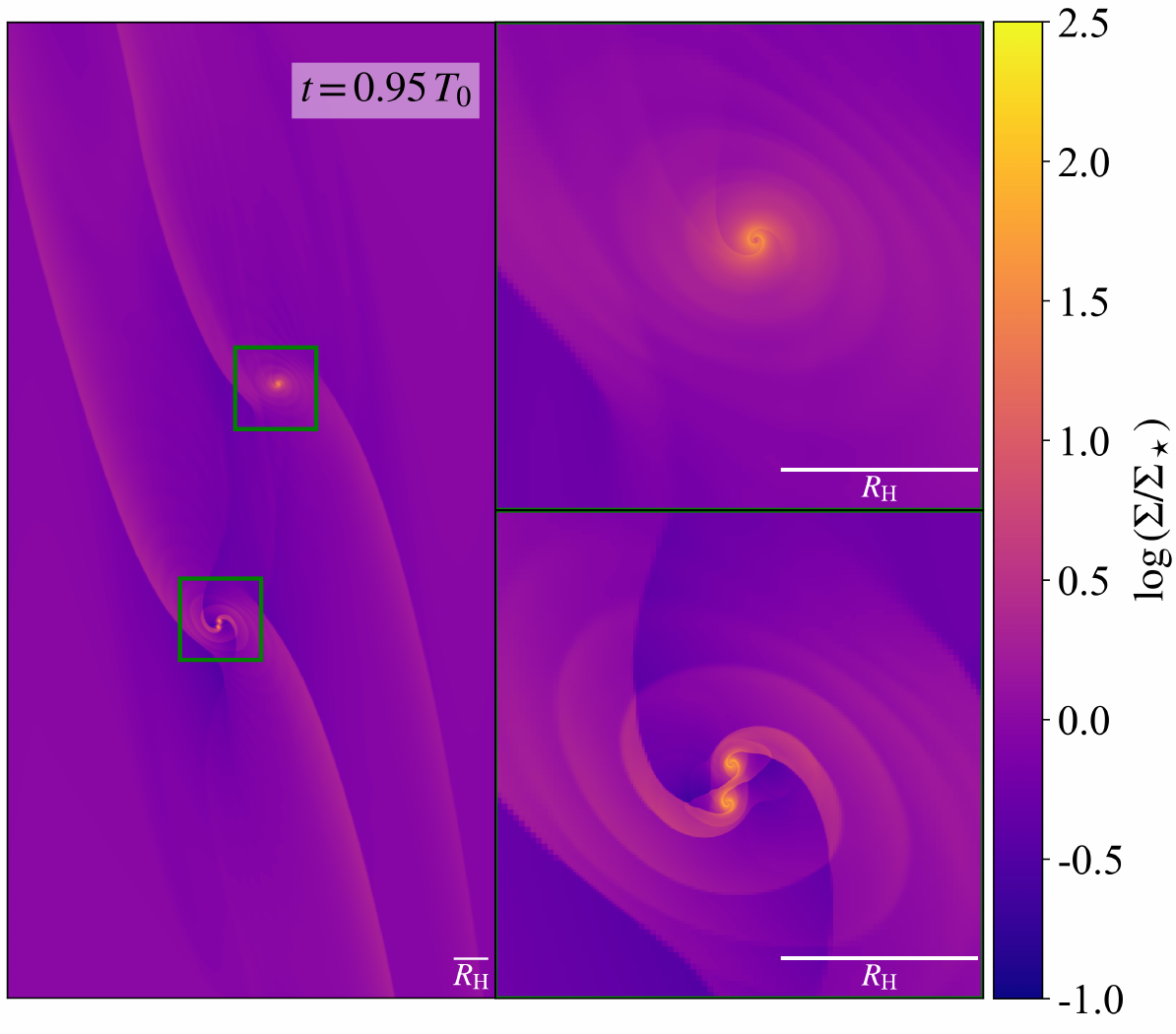}
\caption{The gas density distributions before the interaction between the BBH and the single sBH at $t = 0.95 T_{0}$. The right panels provide zoomed-in views of the regions around the single sBH and the BBH, respectively. There are inner and outer spiral arms on scales of $R_{\rm H}$ for both the single sBH and the BBH. Within the sphere, the single sBH is surrounded by a prograde circumstellar disk\,(CSD), while the BBH is surrounded by a circumbinary disk\,(CBD) composed of two CSDs connected by a spiral arm.
}
\label{fig:before_BSI_den}
\end{figure}

\begin{figure*}
\centering
\includegraphics[scale = 0.65]{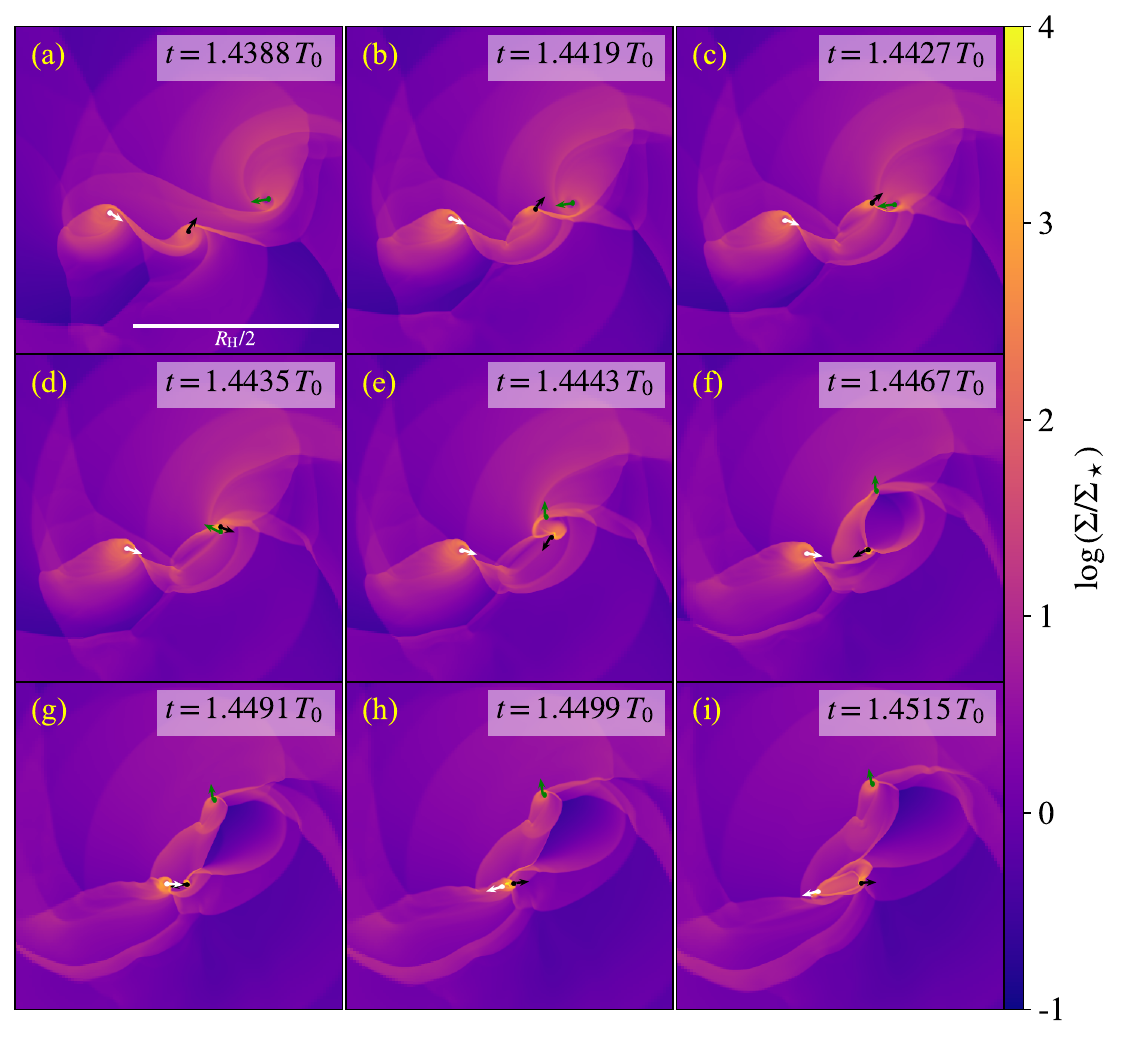}
\caption{Snapshots of the gas surface density at the first close encounter for our fiducial model. The black/white/green dots and arrows represent the position and velocity direction of sBH$_1$/sBH$_2$/sBH$_3$, respectively. As the sBHs approach each other, their CSDs become connected through a new spiral arm. Subsequently, their CSDs collide, forming a high-density region between the two sBHs. As the two sBHs move away from each other, the gas between them will exert a drag force on them.}
\label{fig:under_BSI_den}
\end{figure*}

\subsection{The Initial Conditions and Parameters Setting}
We use the publicly available Eulerian GRMHD code Athena++\,\citep[][]{stone2020ApJS} and couple it with the N-body code REBOUND\,\citep[][]{Rein2012A&A} to simulate the influence of the dense gas on the BSI processes on the AGN disk. Our simulation tracks a 2D shearing box\,\citep{Hawley1995ApJ,Stone1996ApJ} corotating with the AGN disk at the Keplerian angular frequency $\Omega_0$. Similarly to the previous work\,\citep[e.g.,][]{LJR2023ApJ,Dempsey2022ApJ,Rowan2023MNRAS,Whitehead2024MNRAS}, we neglect effects related to gas self-gravity, relativity, radiation, and magnetism, and adopt an isothermal equation of state as a preliminary consideration.

The AGN disk in our model is set to be a Shakura-Sunyaev disk with alpha viscosity $\alpha = 0.1$. The surface density of the disk can be given by: 
\begin{equation}
\begin{split}
\frac{\Sigma} {\Sigma_\star}=\left[\frac{0.1}{\alpha}\right]^{0.8} \left[\frac{M}{10^8\,M_\odot}\right]^{0.2} \left[\frac{\dot{M}}{0.1\dot{M}_{\rm Edd}}\right]^{0.6} \left[\frac{10^3\,R_{\rm g}}{R} \right]^{0.6},
\end{split}
\end{equation}
where $\Sigma_\star = 1.7\times 10^5\,g\cdot \rm cm^{-2}$ and $R_g \equiv GM/c^2$ is the gravitational radius of the SMBH. The equation of state of the disk gas is set to isothermal with $P = \Sigma c_s^2$ with $c_s = H \Omega_0$, where $H$ and $\Omega_0$ are the height of the disk and the angular velocity of the center of the shearing box.  In addition, the disk is geometrically thin with a disk aspect ratio $h\equiv H/R=0.01$. 

The shearing box on the AGN disk is located in a circular Keplerian orbit around the central SMBH at a distance of $R_0 = 10^3\,R_g$. The simulation domain spans $x\in[-7.5R_{\rm H},\,7.5R_{\rm H}]$ and $y\in[15R_{\rm H},\,15R_{\rm H}]$, which is meshed with a base grid of $512 \times 1024$ cells. Based on this grid, we use four levels of Adaptive Mesh Refinement\,(AMR) within $0.2\,R_{\rm H}$ of each sBH, resulting in a total of 5 levels. The cell size within the maximum refinement zone is $\Delta x \sim 0.0016R_{\rm H}$.

In the corotating frame $(x,y)$, the fluid boundary conditions in $x=\pm7.5 R_H$ are set to be the outflow boundary condition, while the refilling boundary conditions are imposed in the boundaries of $y=\pm 15 R_H$\,\citep[e.g.,][]{Whitehead2024MNRAS}. In detail, all ghost cells around $x=\pm7.5 R_H$ are set to the values at the domain edge of the shearing box zone. The ghost cells at $y=-15R_H$ are set to have the values at the domain edge for $x>0$ and at the global disk for $x<0$. In contrast, the ghost cells at $y=15R_H$ the ghost cells are set to have the values at the domain edge for $x<0$ and at the global disk for $x>0$. The physical principle of these value settings is to preserve the shearing feature of the fluid flow at the boundary of the shearing box.

The SMBH mass and the sBHs mass are set to be $M=10^8\,M_\odot$ and $m_1=m_2=m_3=50\,M_\odot$, respectively. The qualitative significance of the gas effect studied in this work will not depend on the choice of the sBH mass. However, considering the growth of sBHs in AGN disks through accretion during the long migration timescale\,($\sim 10^5-10^7\, \rm years\rm $) and possible hierarchical mergers, we select a relatively large mass for the sBHs in this study. Besides, we set the initial semi-major axis \(a_{\rm b0} = R_{\rm H}/5\) of the BBH based on the resolution of our simulation, the gravitational softening length between sBH and gas $\epsilon = 0.08\,a_{\rm b0}$, the sink radius $r_{\rm sink} = 0.8\epsilon$ and the gas removal rate $n_{\rm b}=2.0$. In this work, we adopt the Newtonian gravity with a gravitational softening length $10^{-6}\,R_{\rm H}$ between sBHs in the N-body simulations.


For the BSI process, we use the impact parameter $b$ to describe the radial\,($x$-direction) separation between the BBH and the single sBH, shown in Figure\,\ref{fig:sch_rep}. The angular\,($y$-direction) separation is set as $\Delta \phi = 25R_{\rm H}/R_0$. Thus, the initial positions of the three sBHs in the corotating frame are $\left(-b/2-a_{\rm b0}/2, -12.5R_{\rm H}\right)$, $(-b/2+a_{\rm b0}/2, -12.5R_{\rm H})$ and $(b/2, 12.5R_{\rm H})$, respectively. The center of mass of the BBH and the single sBH both perform Keplerian motion relative to the central SMBH. To highlight the impact of gas on the BSI process, in our simulation, we turn off gas gravitational effects on the sBHs until the distance between the BBH's center of mass and the single sBH is less than $2\,R_{\rm H}$.  Physically this is because before the binary moves close to the incoming sBH,  the sBH and its accretion gas background comove and follow an independent Keplerian motion around the SMBH from that of the BBH.  In this way, our hydrodynamical simulations have reached a steady state before the BSI process begins\,(as shown in Figure\,\ref{fig:before_BSI_den}).  In this work, we explore the parameter space with the surface density of the AGN disk \(\Sigma_0 \in [0,\,5]\,\Sigma_\star\) and the impact parameter \(b \in [1.6,\,2.5]\,R_{\rm H}\).

\section{FIDUCIAL RESULTS} \label{sec:Fiducial_results}
In this section, we present the simulation results under the fiducial model, in which we set a uniform ambient surface density $\Sigma_0=\Sigma_\star$ and the impact parameter $b=2.005\,R_{\rm H}$.
This fiducial model will provide a detailed dynamic analysis of the impact of dense gas on the BSI process.

\subsection{Gas Morphology}
In our model, the dense gas interacts with the three sBHs through gravitational forces as described by equation\,\eqref{eq:a_gas}. Therefore, the gas morphology is essential for determining the $dm_{ij}$ in equation\,\eqref{eq:a_gas}.
Figure\,\ref{fig:before_BSI_den} shows the structural morphology of the dense gas around the three sBHs before the BSI at $t= 0.95T_0$, where $T_0=2\pi/\Omega_0$ is the orbital period of the shearing box. On large scales\,($>R_{\rm H}$), the gas spiral arms caused by the single sBH and the BBH are morphologically similar, with clear inner leading and outer trailing spiral arms that extend from the $R_{\rm H}$ away from the sBHs. Within the Hill sphere, gas accumulates around the single sBH, forming a prograde circum-single disk\,(CSD). Around the binary, the two prograde CSDs are connected by a spiral arm, forming a larger prograde circumbinary disk\,(CBD). Notably, the accretion of the gas by the sBH creates a low-density region that appears within the sink radius. 

Figure\,\ref{fig:under_BSI_den} displays the evolution of gas morphology during the BSI process, where only the gas morphology at the first close encounter of the three sBHs is shown. The black/white/green dots and arrows represent the positions and velocity directions of the three sBHs, respectively. As the two sBHs gradually approach each other, new spiral arms form between their CSDs. When they get close enough, their CSDs collide, forming a high-density gas region between the two sBHs. The dense gas in this region exerts a drag force on the two sBHs when they separate, leading to energy dissipation from the three-body system to the gas\,(see Section\,\ref{sec:energy_dissipation}).

\begin{figure}
\centering
\includegraphics[scale=0.48]{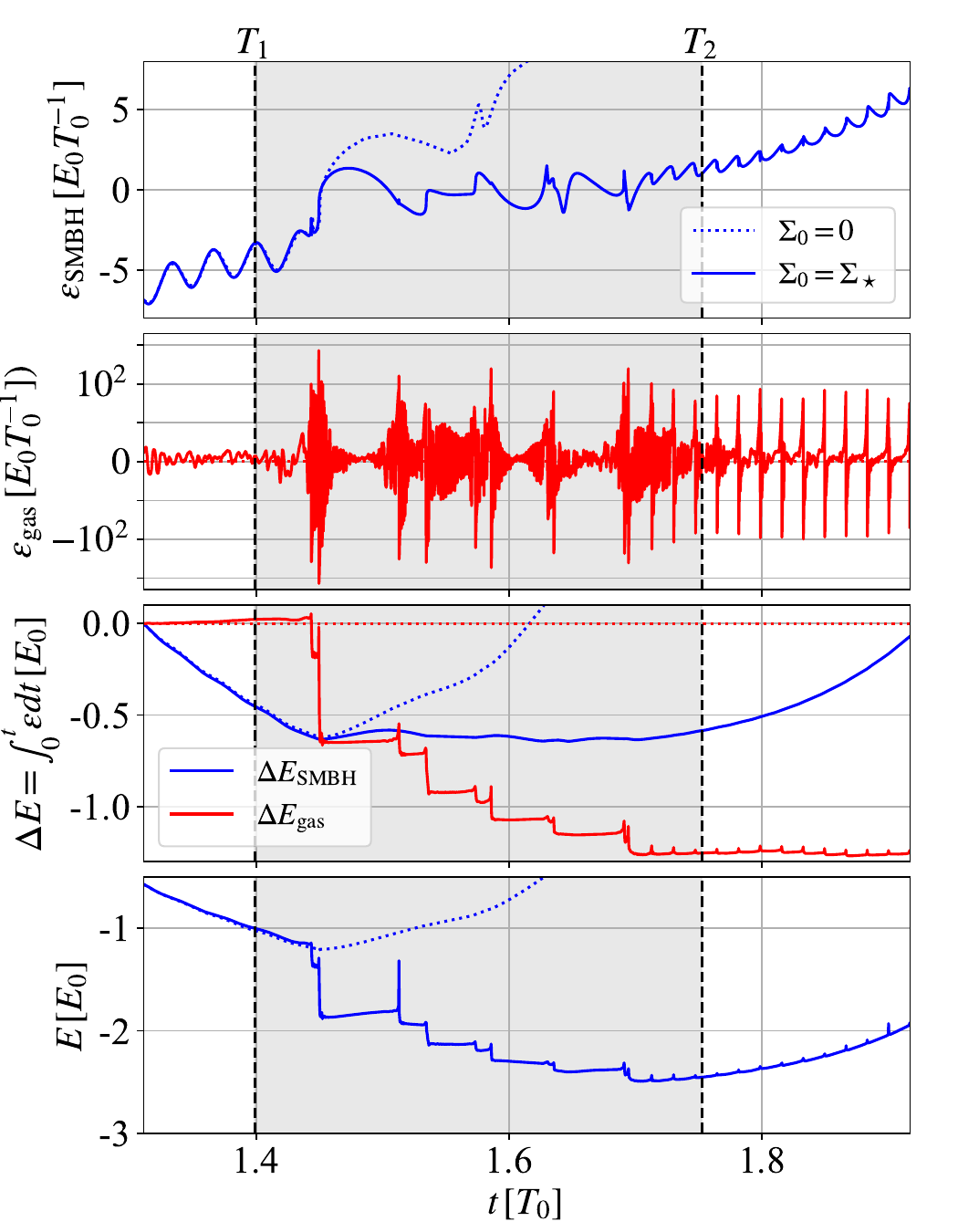}
\caption{The energy dissipation of the three-body system in our fiducial model. The two vertical dashed lines represent the start\,($T_1$) and end times\,($T_2$) of the BSI, determined by whether the maximum distance between the three sBHs exceeds $R_{\rm H}$.
The top two panels show the energy change rate\,(power) due to the SMBH and the gas in the unit of $E_0T_0^{-1}$, respectively, for $\Sigma_0 = 0$\,(dotted line) and $\Sigma_0 = \Sigma_\star$\,(solid line), where $E_0 = |E(T_1)|$ is the absolute value of the initial energy of the three-body system at $T_1$. The bottom two panels show the time integration of the power due to the SMBH and the gas, as well as the energy evolution of the three-body system.  }
\label{fig:energy_evolution}
\end{figure}

\begin{figure}
\centering
\includegraphics[scale=0.48]{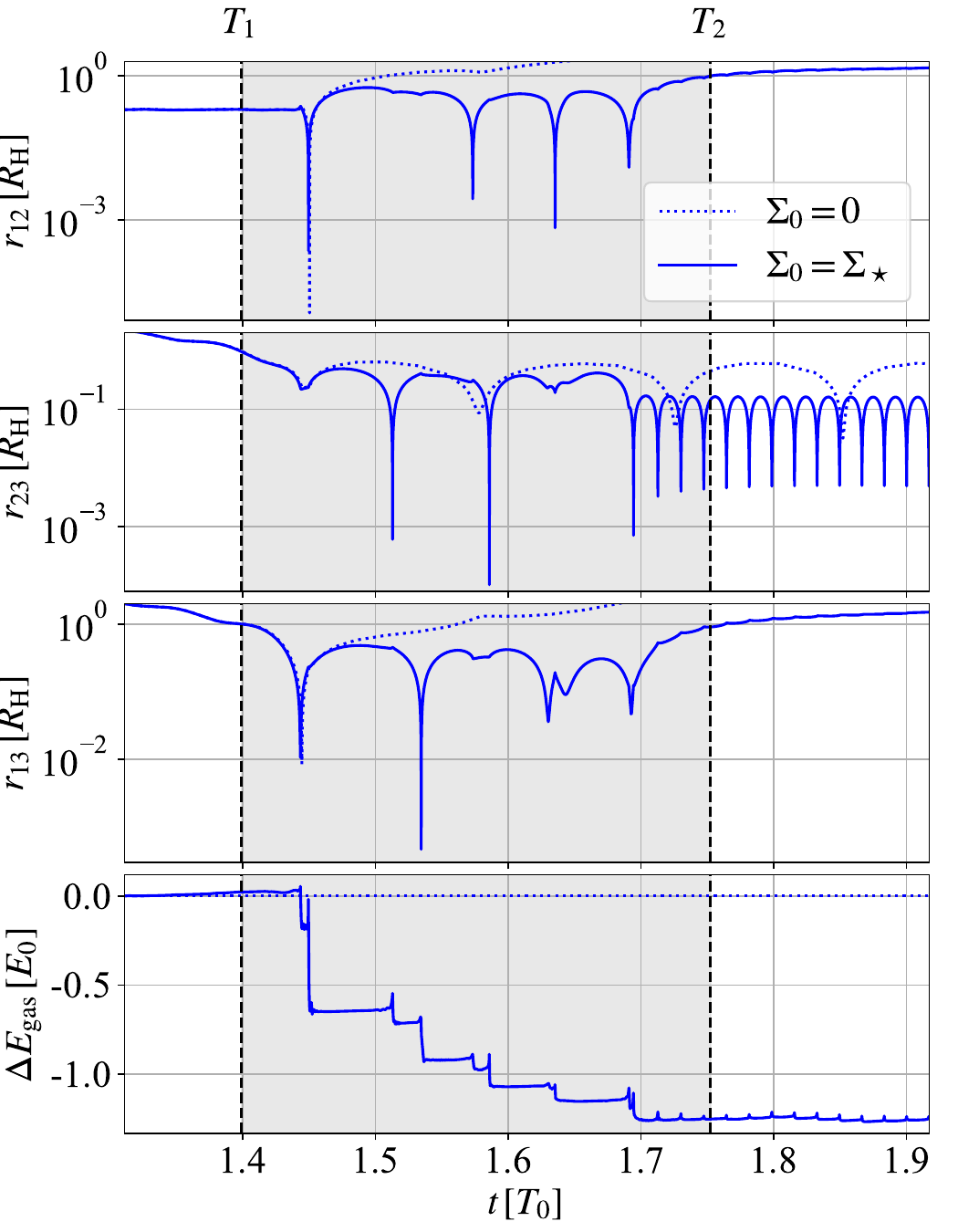}
\caption{The evolution of the distance between two sBHs in the three-body system. The dotted and solid lines represent the conditions with initial surface densities of $\Sigma_0 = 0$ and $\Sigma_0 =\Sigma_\star$, respectively. Consistent with Figure\,\ref{fig:energy_evolution}, the bottom panel shows the time integration of the power by the gas. The pulsed energy losses of the three-body system correspond to moments of close encounters between the sBHs.}
\label{fig:distance}
\end{figure}

\subsection{Energy Dissipation during the BSI Process} \label{sec:energy_dissipation}
To quantitatively describe the role of gas in the three-body BSI process, this subsection investigates the energy exchange between the three sBHs system and its environment, i.e., the SMBH and the gas. Hereafter, we denote the energy change rate of the three sBHs system\,(power) due to the gravitational influence of the SMBH as $\varepsilon_{\rm SMBH}$, and that caused by the disk gas as $\varepsilon_{\rm gas}$. The energy dissipation of the three-body system, defined in \cite{Rowan2023MNRAS,Whitehead2024MNRAS}, can be determined by this power.
\\

The power can be calculated by taking the derivative of the total energy with respect to time: 
\begin{equation}\label{eq:power}
\begin{split}
   & \varepsilon \equiv \frac{{\rm d}E}{{\rm d} t} = \varepsilon_{\rm SMBH} + \varepsilon_{\rm gas},\\
   & \varepsilon_{\rm SMBH} = \sum_{1\leq k<l\leq 3}^3 \frac{m_km_l}{M_{\rm sBH}} (\bm{v}_k-\bm{v}_l)\cdot (\bm{a}_{k,\rm SMBH}-\bm{a}_{l,\rm SMBH}),\\
    &    \varepsilon_{\rm gas} = \sum_{1\leq k<l\leq 3}^3 \frac{m_km_l}{M_{\rm sBH}} (\bm{v}_k-\bm{v}_l)\cdot (\bm{a}_{k,\rm gas}-\bm{a}_{l,\rm gas}).
 \end{split}
\end{equation}
Here, the acceleration of the three sBHs $k=(1,\,2,\,3)$ due to the gravitational force of SMBH is
\begin{equation} \label{eq:a_SMBH}
    \bm{a}_{k,\rm SMBH} = GM\frac{\bm{r}_k-\bm{r}_{\rm SMBH}}{|\bm{r}_k-\bm{r}_{\rm SMBH}|^3},
\end{equation}
while the acceleration due to the gas is given by equation\,\eqref{eq:a_gas}. 

\begin{figure} 
\centering
\includegraphics[scale=0.4]{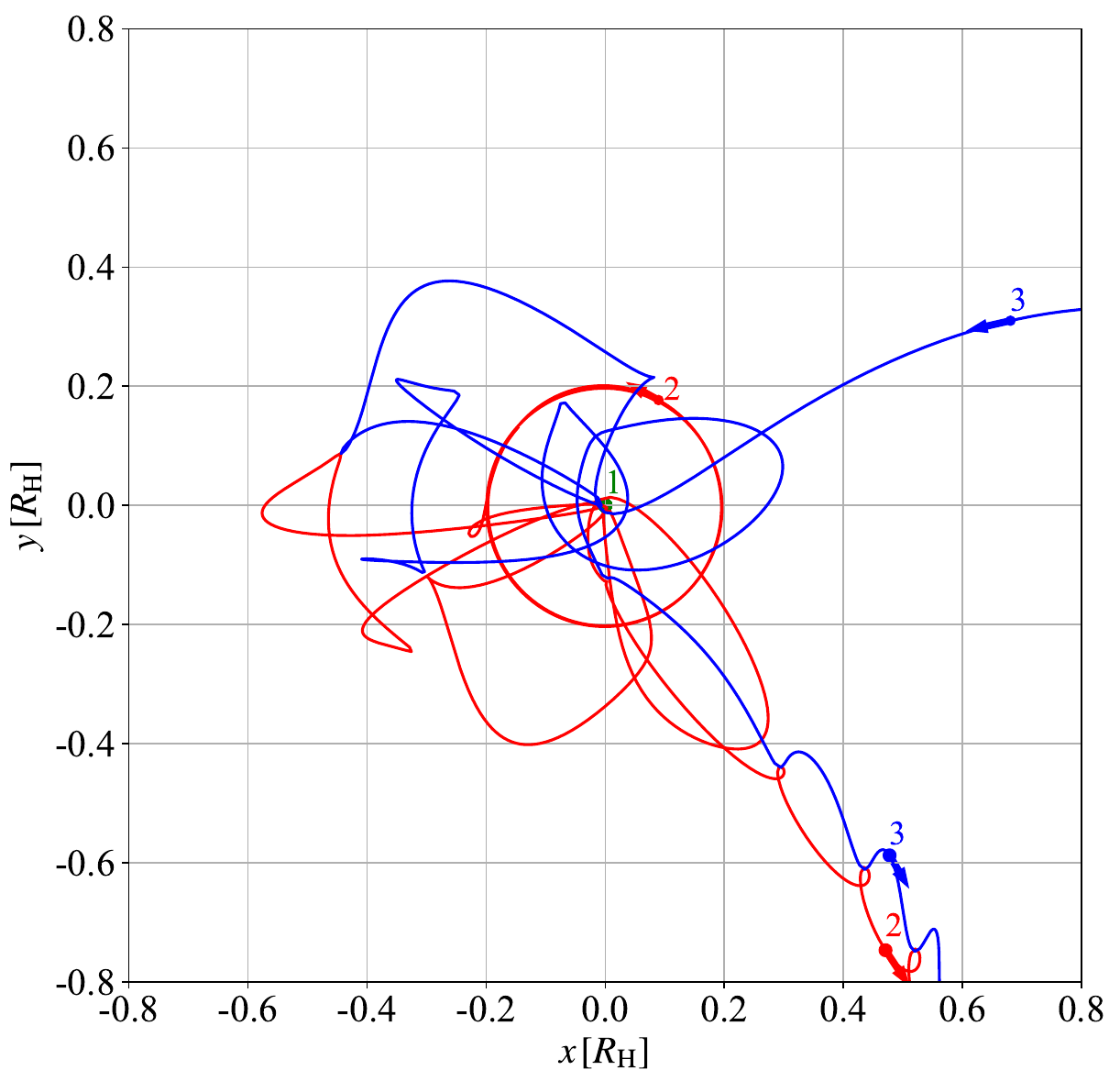}
\caption{Configurational trajectory evolution of the three-body system, where the green, red, and blue dots symbolize sBH$_{1}$/sBH$_{2}$/sBH$_{3}$, respectively. We have chosen to analyze this configuration using \emph{the reference frame of sBH$_{1}$}, which initially pairs up with sBH$_{2}$
to form a binary system. After BSI, sBH$_{3}$ eventually takes the place of sBH$_{1}$ in the end-state BBH, while sBH$_{1}$ is expelled from the Hill sphere.}
\label{fig:trajectory}
\end{figure}

\begin{figure*}
\centering
\includegraphics[scale = 0.9]{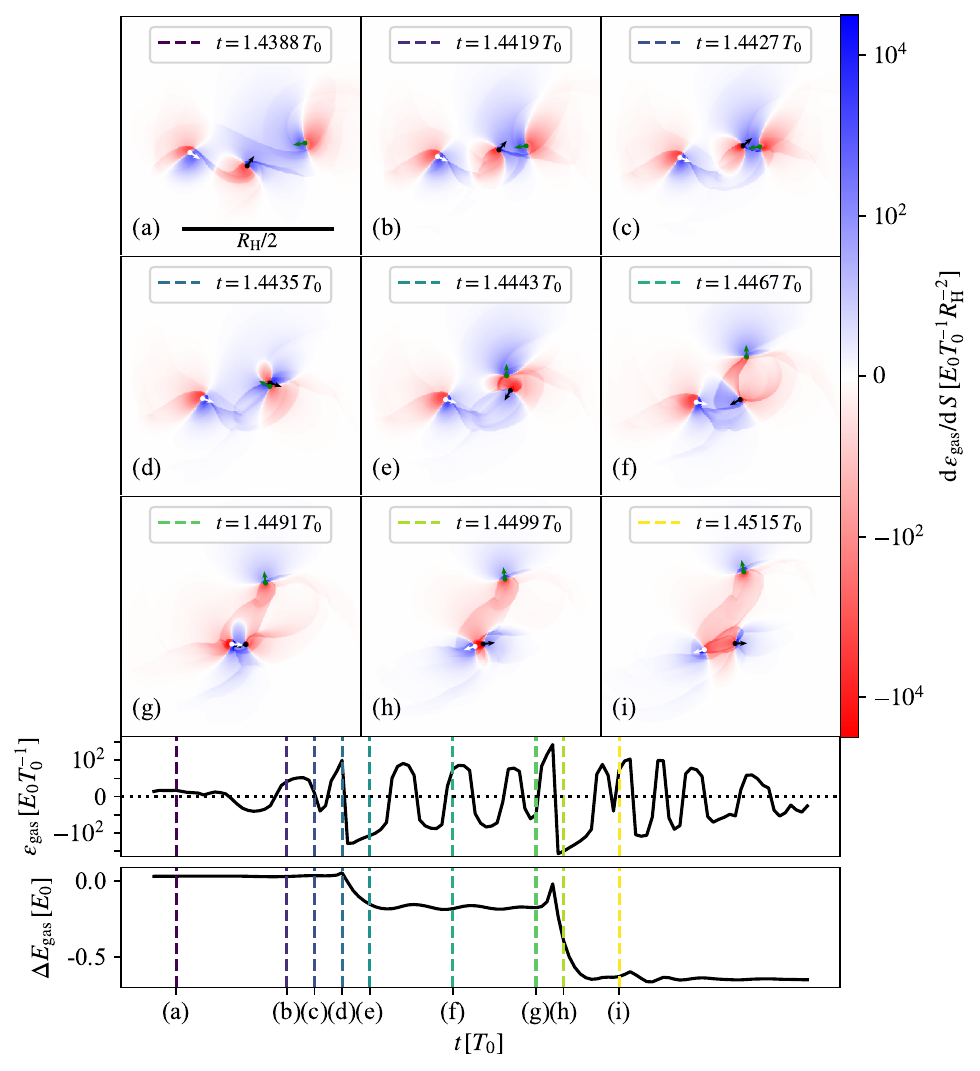}
\caption{The energy changes rate\,(power) per unit area of the three-body system by disk gas for our fiducial model. The black/white/green dots and arrows represent the position and velocity direction of sBH$_1$/sBH$_2$/sBH$_3$, respectively. The red regions represent the energy lost by the three-body system to the gas, while the blue regions represent the energy gained by the three-body system from the gas. Consistent with Figure\,\ref{fig:energy_evolution}, the bottom two panels illustrate the power and the time integration of the power. After the sBHs' CSDs collide, the gas motion lags behind the sBHs. As the sBHs move apart, the net effect of this process is an energy loss for the three-body system.  }
\label{fig:force_under_BSI}
\end{figure*}

Using equations\,\eqref{eq:a_gas}, \eqref{eq:total_EL}, \eqref{eq:power} and \eqref{eq:a_SMBH},  we show the energy evolution of the three-body system along with the work done by the SMBH and gas during the BSI process in Figure\,\ref{fig:energy_evolution}, and the evolution of the distance between different sBHs $r_{ij}$ in the three-body system in Figure\,\ref{fig:distance}.  The dotted and solid lines show the energy evolution at $\Sigma = 0$\,(without gas) and $\Sigma = \Sigma_\star$, respectively.   The evolution of the three-body configuration is shown in Figure\,\ref{fig:trajectory} where the green/red/blue spot represents the sBH$_{1/2/3}$. We choose to describe the configuration in the reference frame of the sBH$_{1}$, which forms the initial binary with the sBH$_{2}$. As shown in Figure\,\ref{fig:trajectory}, the sBH$_{3}$ replaces the  sBH$_{1}$ in the end-state binary, while the  sBH$_{1}$ is rejected out of the Hill sphere.   Furthermore, to better illustrate the dynamical process, we plot the spatial distribution of the contribution of the work done by the gas gravity, defined as the power per unit area ${\rm d} \,\varepsilon_{\rm gas}/ {\rm d} \,S $, during the first close encounter in Figure\,\ref{fig:force_under_BSI}. 

The full process can be divided into three stages by the two vertical dashed lines $t=T_1$ and $t= T_2$ in Figure\,\ref{fig:energy_evolution}. The $T_{1}$ and $T_2$ are defined as the start and end time of the BSI process, determined by whether the maximum distance between the three sBHs exceeds $R_{\rm H}$. 

$\bullet$ When $t<T_1$, the distance between the incoming sBH and the BBH center of mass decreases, leading to a reduction in the negative power generated by the tidal force along their relative distance. This accounts for why the power of the three-body system due to the SMBH is negative and why $|\varepsilon_{\rm SMBH}|$ shows a decreasing trend. Furthermore, the incoming sBH is outside the Hill sphere of the BBH, and the distance between the BBH components undergoes periodic change due to the gravitational tidal force of the central SMBH. This further explains why there is an oscillatory feature in the $\varepsilon_{\rm SMBH}$.  

Moreover, as shown in Figure\,\ref{fig:force_under_BSI}, when the distance between the sBHs is relatively large, the power contributed by the gas gravity in front of and behind the sBHs nearly cancels out, resulting in the $\varepsilon_{\rm gas}$ almost approaching zero. Therefore, during this stage, the dense gas plays a minor role so that the evolution almost follows the case when $\Sigma_0=0$.

$\bullet$ When $t>T_2$, the three sBHs finish their complicated interactions, and an end state is formed so that one of the sBH flies away from the Hills sphere around the residue BBH, which means the distance between the flying-away sBH and the residue BBH's center of mass is increasing, together with the increasing of the SMBH gravitational tidal force. In this case, the tidal force of the SMBH does positive work on the three-body system and the power $\varepsilon_{\rm SMBH}$ is increasing. Furthermore, the gravitational tidal force affects the orbital distance of the BBH components, which also exhibits an oscillatory feature. 

Importantly, our simulation reveals that the residue BBH has a highly prograde eccentric and significantly more compact orbit compared to the initial BBH\,(before BSI: $a_{\rm b0} = 0.2\,R_{\rm H}$, $e_{\rm b0} = 0$, after BSI: $a_{\rm b0} = 0.08\,R_{\rm H}$, $e_{\rm b0} = 0.95$ for $\Sigma_0 = \Sigma_\star$). For such a highly eccentric BBH, the power exerted by the gas (\(\varepsilon_{\rm gas}\), as shown in the second panel of Figure\,\ref{fig:energy_evolution}) exhibits an impulsive pattern of increases and decreases, in contrast to the initial circular BBH. The gas-induced power is positive as the binary moves from apocenter to pericenter, while it is negative from pericenter to apocenter.
Our analysis is consistent with the simulation results of \cite{Calcino2024ApJ} for eccentric binaries.

$\bullet$ The BSI happens during $T_1<t<T_2$, where the three sBHs undergoes complicated gravitational encounter process.  Since these three sBHs are close to each other with the typical distance $\leq R_{\rm H}$, the work done by the SMBH is thus smaller compared to the stages $t<T_1$ and $t>T_2$.
During each encounter when the two sBHs get close enough, their CSDs collide, and a large amount of gas accumulates between them. The accumulated gas forms a high-density region that gravitationally attracts these two encountering sBHs, and the gas does positive work on the two sBHs so that the energy of the two sBHs from the gas increases. In our fiducial model, this process corresponds to the process from (c) to (d) in Figure\,\ref{fig:force_under_BSI} and is quantitatively shown in the bottom two panels of Figure\,\ref{fig:force_under_BSI}.   

After the two sBHs with high relative velocity fly by each other, their separation will be gravitationally dragged by the accumulated dense gas corresponding to negative work.  In this way, the close encounter processes result in significant energy dissipation in the three-body system, which corresponds to the process from (d) to (e) in Figure\,\ref{fig:force_under_BSI}.

Moreover, the multiple encounter events among these three sBHs happen intermittently, and so do the corresponding energy dissipations to the gas. For a better illustration, we integrate the power of the gas and SMBH over time, shown in the third panel of Figure\,\ref{fig:energy_evolution}. It is evident from the figure that during the BSI process, the work done by the gas shows several intermittent steep declines\,(corresponds to the process from (d) to (e) in Figure\,\ref{fig:force_under_BSI}), while the work done by the SMBH almost stays constant. These significant declines correspond to the intermittent encounter events, which can be seen from the evolution of the distance $r_{ij}$ between the $i_{\rm th}$ sBH and the $j_{\rm th}$ sBH in Figure\,\ref{fig:distance}.  

In summary, the impact of dense gas on the BSI process in our fiducial model is characterized by the following three features. (1) Compared to the gas-free case, the gas also increases the duration of the three-body interactions featured by intermittent multiple close encounters;  (2) During the intermittent multiple close encounters, as shown in the bottom panel of Figure\,\ref{fig:energy_evolution}, the presence of the gas leads to a decrease in the total energy of the three-body system; (3) The gas assisted BSI process hardens the end-state BBH system\,($a_{\rm b} = 0.32\,R_{\rm H}$, $e_{\rm b} = 0.77$ for $\Sigma_0 = 0$ and $a_{\rm b} = 0.08\,R_{\rm H}$, $e_{\rm b} = 0.95$ for $\Sigma_0 = \Sigma_\star$), as demonstrated in $r_{23}$-panel of  Figure\,\ref{fig:distance}.

\begin{figure*} 
\centering
\includegraphics[scale=0.9]{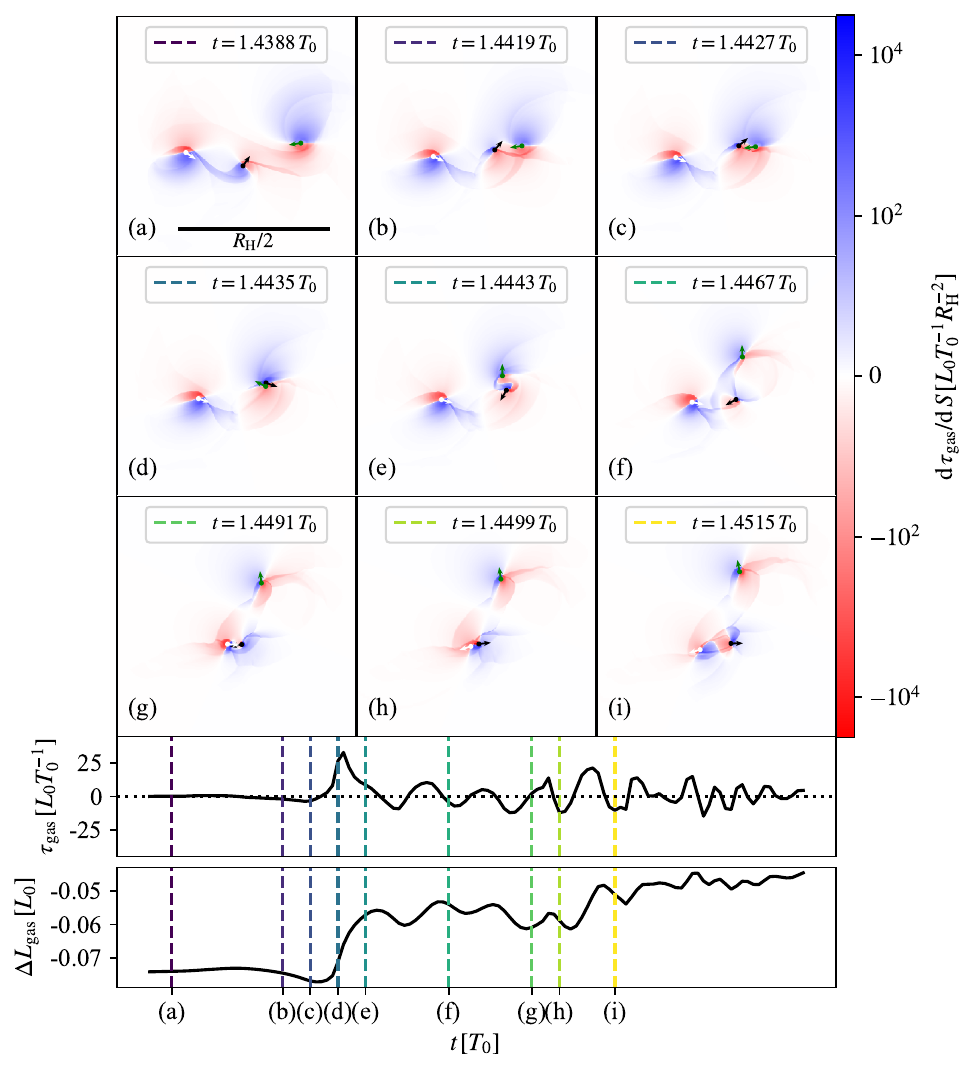}
\caption{ The torque per unit area of the three-body system by disk gas for our fiducial model in the unit of $L_0T_0^{-1}R_{\rm H}^{-2}$, where $L_0 = |L(T_1)|$ is the absolute value of the initial angular momentum of the three-body system at $T_1$. Red regions indicate the angular momentum lost by the three-body system to the gas, while blue regions represent the angular momentum gained by the three-body system from the gas. The bottom two panels display the torque and the time integration of the torque, respectively. }
\label{fig:torque}
\end{figure*}

\begin{figure} 
\centering
\includegraphics[scale=0.42]{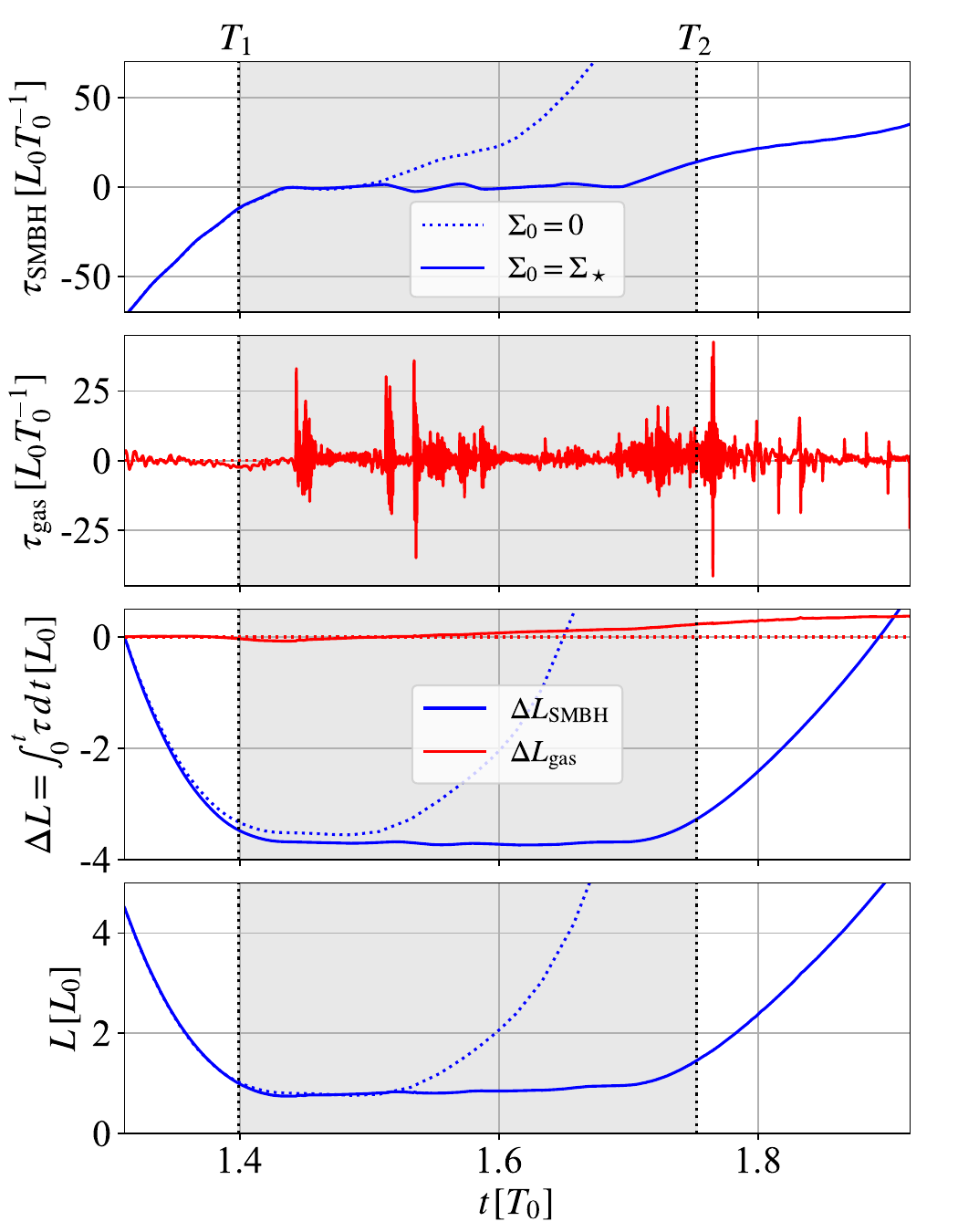}
\caption{The angular momentum dissipation of the three-body system in our fiducial model. The top two panels  show the angular momentum change rate\,(torque) due to the SMBH and the gas in the unit of $L_0T_0^{-1}$, respectively, for $\Sigma_0 = 0$\,(dotted line) and $\Sigma_0 = \Sigma_\star$\,(solid line). The bottom two panels show the time integration of the torque due to the SMBH and the gas, as well as the angular momentum evolution of the three-body system, respectively.}
\label{fig:AG_evolution}
\end{figure}

\subsection{Angular Momentum Dissipation}
Based on the Equation\,\eqref{eq:total_EL}, the torque generated by both the SMBH and the gas can be derived as follows:
\begin{equation}
\begin{split}
&    \tau=\frac{{\rm d}\, L}{{\rm d}\,t} = \tau_{\rm SMBH}+ \tau_{\rm gas},\\
&    \tau_{\rm SMBH}= \sum_{k=1}^3 m_k (\bm{r}_k - \bm{r}_c) \times (\bm{a}_{k,\rm SMBH}- \bm{a}_c),\\
&    \tau_{\rm gas} = \sum_{k=1}^3 m_k (\bm{r}_k - \bm{r}_c) \times (\bm{a}_{k,\rm gas}- \bm{a}_c).
    \end{split}
\end{equation}
Using the above formula, we can also calculate the distribution of the torque per unit area acting on the three-body system contributed by the gas gravity in Figure\,\ref{fig:torque}. The time evolution of the torque and the corresponding change of total angular momentum of the three-body system is shown in the bottom two panels of Figure\,\ref{fig:torque} and at a longer timescale in Figure\,\ref{fig:AG_evolution}. The top two panels of Figure\,\ref{fig:AG_evolution} show the torque of the SMBH's tidal forces $\tau_{\rm SMBH}$ and that of the gas gravitational forces $\tau_{\rm gas}$.  The simulation shows that the torque contributed by gas gravity oscillates around the zero value and thereby can not accumulate and significantly affect the angular momentum. In the lower two panels, we show the time integral of the torque and the evolution of angular momentum. When the CSDs of two sBHs collide, the gas accumulated between them exerts a drag force along the line connecting the two sBHs as they move apart. The tangential component of this drag force relative to the line connecting the sBHs is relatively small, resulting in a minimal impact of the gas on the total angular momentum of the three-body system during the BSI process. Thus, different from energy dissipation, the torque during the approach and separation of the binary and the single sBH is predominantly governed by the central SMBH.

\begin{figure*}
\gridline{\fig{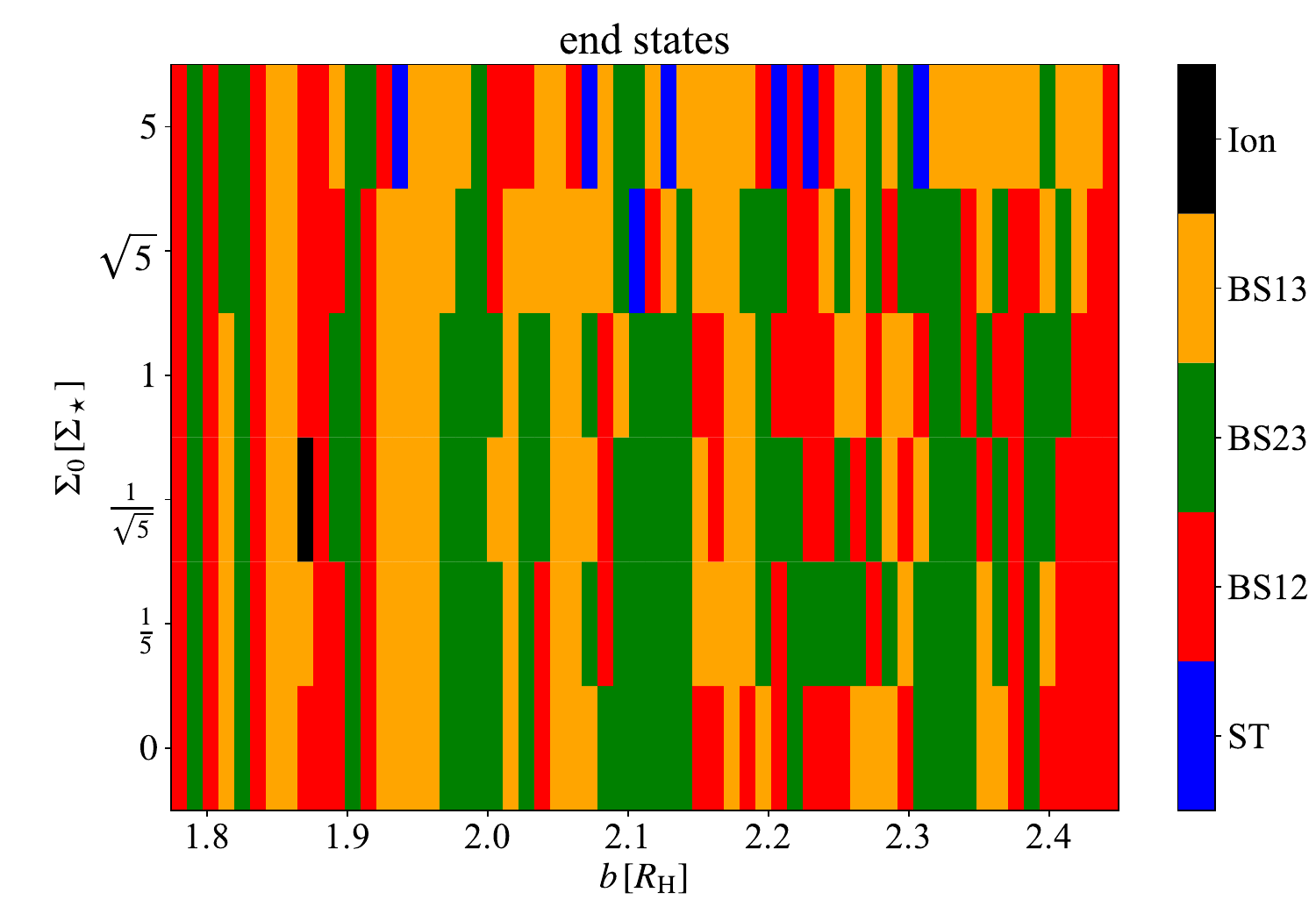}{0.5\textwidth}{}
          \fig{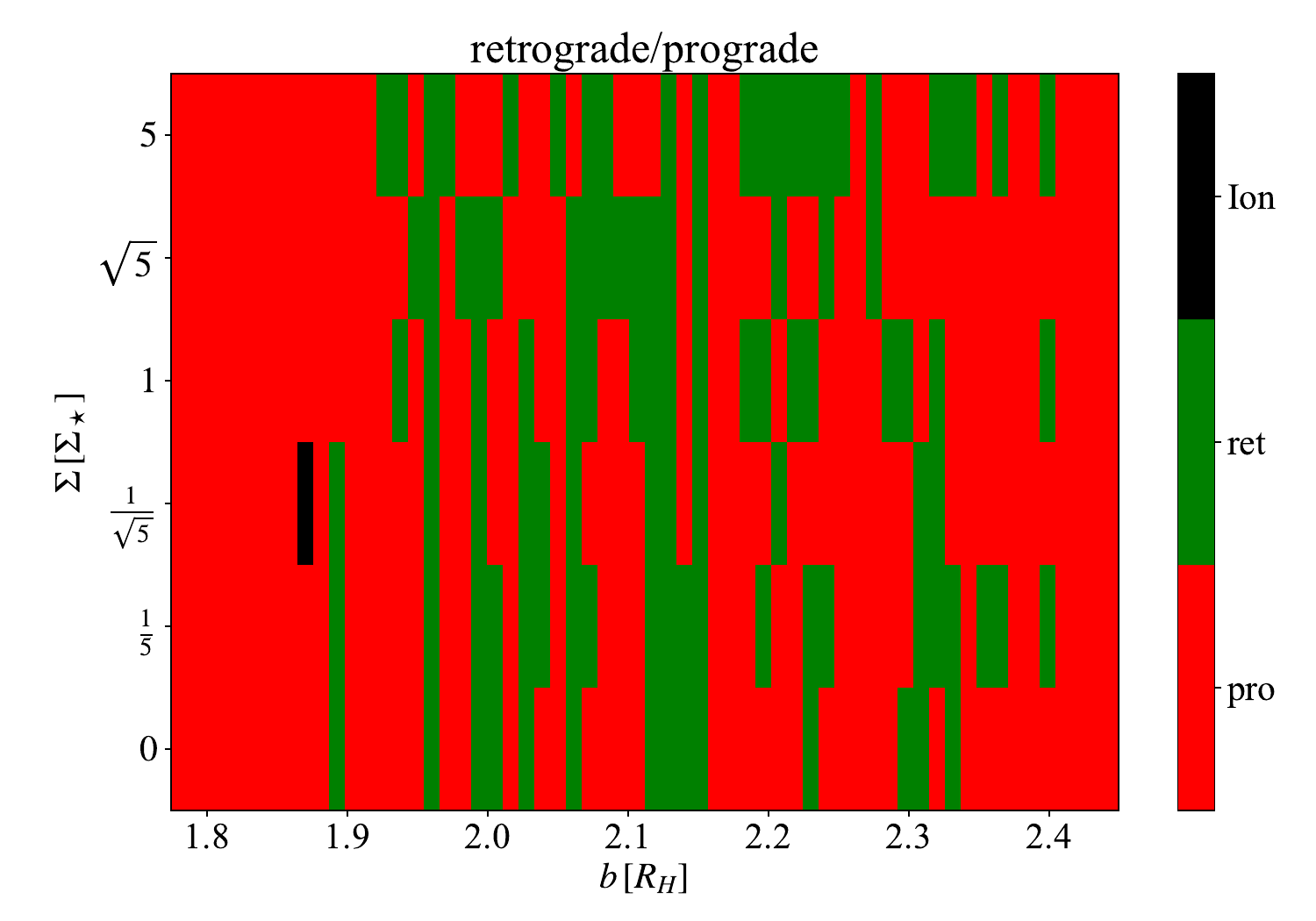}{0.5\textwidth}{}
          }
\caption{The end states of BSI for our 360 simulations. In the left panel, we divide the end states into ST\,(blue), BS12\,(red), BS23\,(green), BS13\,(orange) and Ion\,(black). In the right panel, we show the spin directions of the end-state binaries. Prograde and retrograde configurations are represented in red and green, respectively. The black area represents the formation of an ionization state, where no bound binary exists after the BSI process.}
\label{fig:end_state}
\end{figure*}

\begin{figure*}
\gridline{\fig{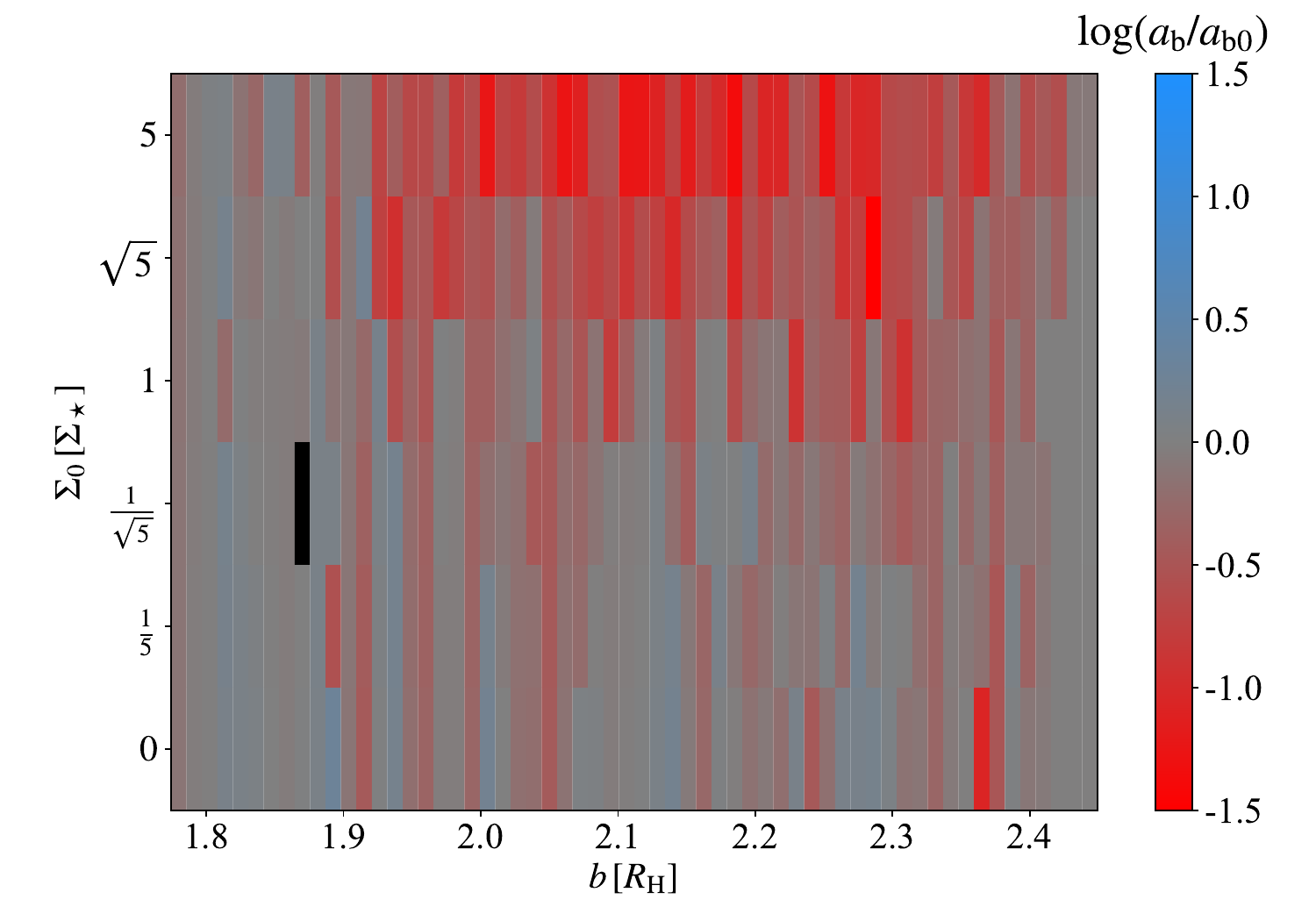}{0.5\textwidth}{}
          \fig{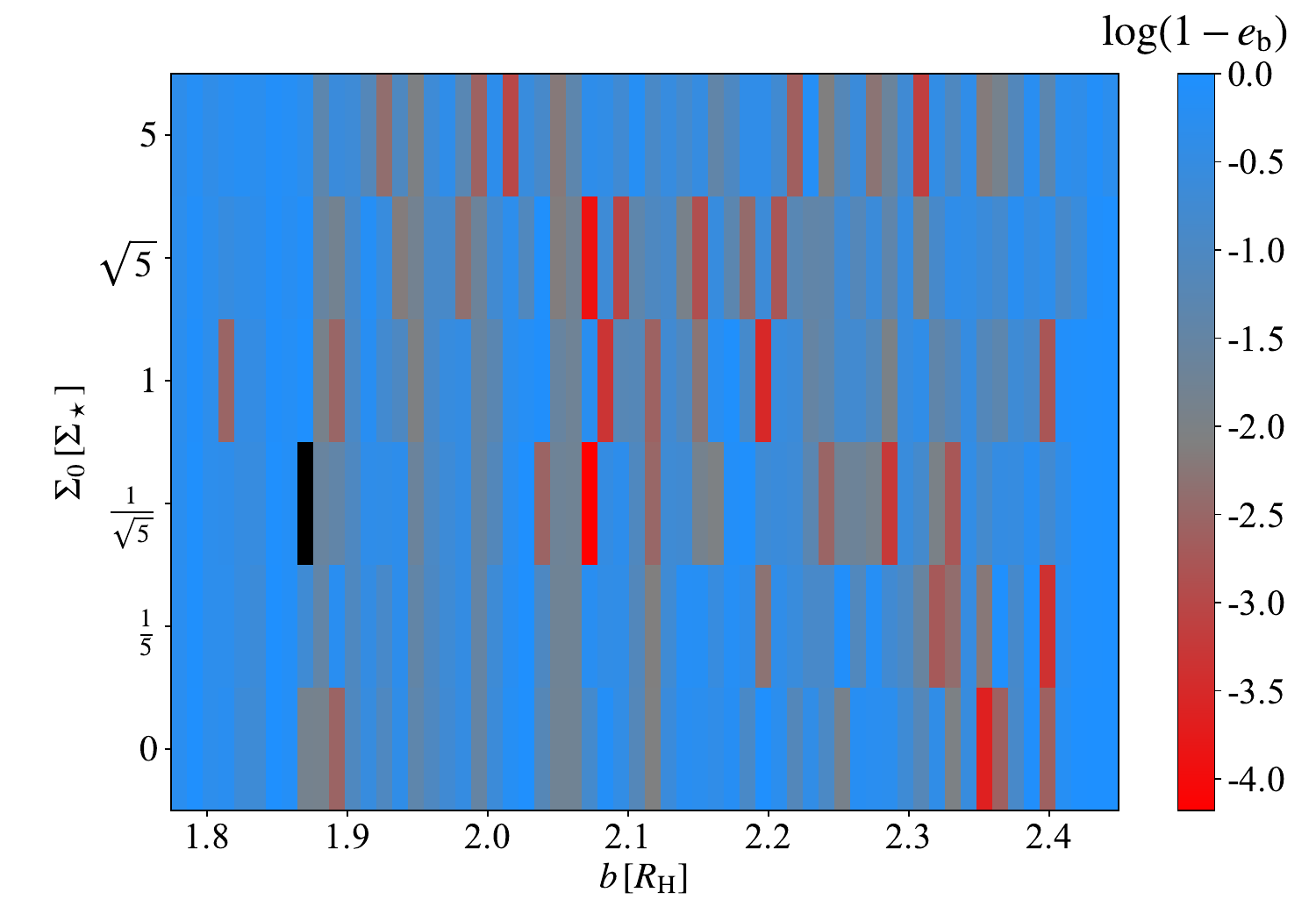}{0.5\textwidth}{}
          }
\caption{The orbital parameters distribution of the end-state binaries in our simulations. In the left panel, we show the ratio between the semi-major axis of the end-state binary and that of the initial binary. As the density increases, this ratio gradually decreases, indicating that the presence of gas makes the end-state binary more compact. In the right panel, we show the distribution of $1-e_{\rm b}$ in the log scale. Although the presence of gas alters the eccentricity, it does not significantly change the overall distribution of eccentricity.  The black area represents the formation of an ionization state.}
\label{fig:af_ef}
\end{figure*}

\begin{figure*} 
\centering
\includegraphics[scale=0.5]{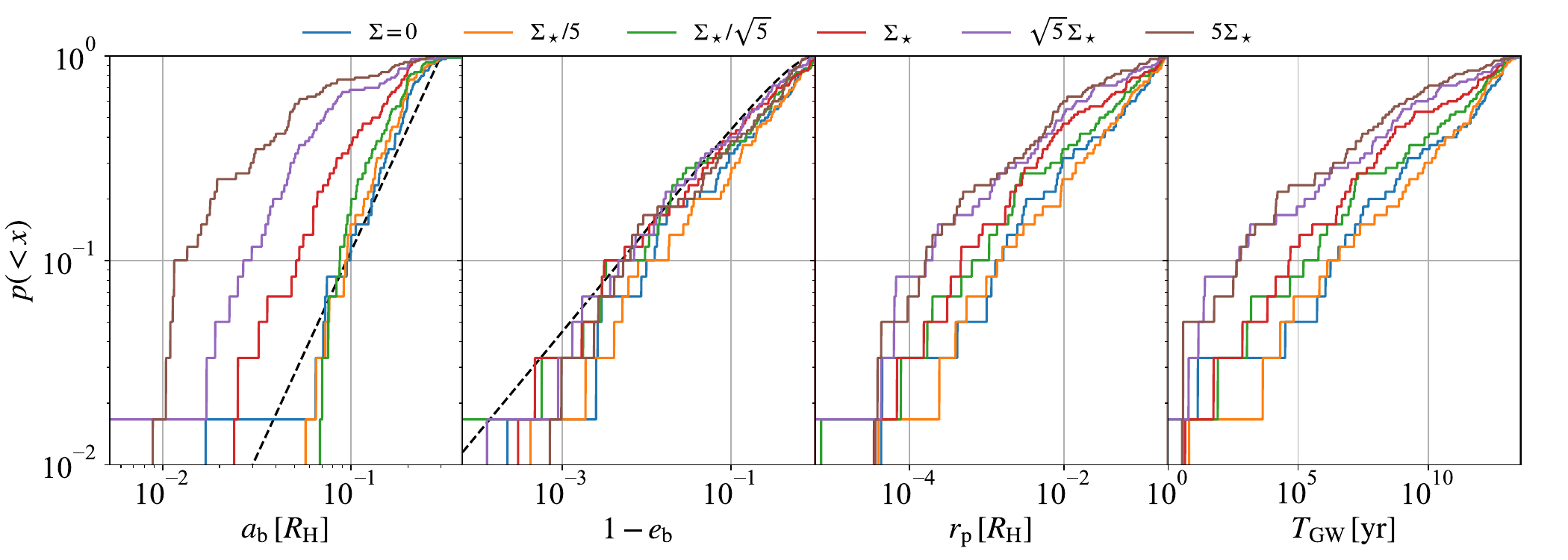}
\caption{ The CDFs of end-state BBH orbital parameters at various densities are shown across the four panels: from left to right, they display the CDF of the final semi-major axis $a_{\rm b}$, eccentricity $1 - e_{\rm b}$, pericenter $r_{\rm p}$ and GW merger time $T_{\rm GW}$. The black dashed lines in the first two panels represent the theoretically predicted distributions of the semi-major axis and eccentricity, which is given by equation\,\eqref{eq:CDF}.
The presence of gas makes the end-state-BBH semi-major axis more compact, leading to a shorter gravitational wave merger timescale, while it has little effect on the CDF of eccentricity.}
\label{fig:aeT_dis}
\end{figure*}

\section{Statistical Research of Parameter Space} \label{sec:parameter_space}
In this section, we study the influence of the gas presence on the statistics of the end states of the BSI interactions. We simulate the BSI process by scanning the two-dimensional parameter space of the density $\Sigma_0$ and impact parameter $b$. This parameter space is divided into $6\times 60=360$ grids in the following way: the 60 impact parameters are chosen to be as $b = b_{\rm min} + i \times (b_{\rm max}- b_{\rm min})/n$, where $b_{\rm min}=1.6\,R_{\rm H}$, $b_{\rm max} =2.5\,R_{\rm H} $, $n=80$ and $i\in [16,\,75]$ is an integer; the 6 density values are $(0,\,\,1/5,\,1/\sqrt{5},\,1,\,\sqrt{5},\,5)\,\Sigma_\star$. For each $(\Sigma_0,b)$, our simulations run for a total $3\,T_0$ duration. The end state is defined at the time\,($T_2$) when the distance between two sBHs in the three-body system remains greater than $R_{\rm H}$ until the termination of simulation at $3\,T_0$. 

\subsection{Gas-influenced End States}
The end states of a BSI process can be categorized into three types: stable triple\,(ST), binary + single sBH\,(denoted as BS12/BS23/BS13, which means sBH$_1$/sBH$_2$/sBH$_3$ and sBH$_2$/sBH$_3$/sBH$_1$ forms a residue binary), and ionization\,(Ion). For the ST state, all three sBHs remain within the Hill sphere, consisting of a compact binary and a third body that orbits the center of mass of the binary along an elliptical trajectory. The end-state moment of ST state is defined as the first time after the formation of an ST system when the third sBH and the BBH are at their apocenter. 

Previously, \cite{Fabj2024arXiv} argued that gas primarily affects BSI processes involving long-duration interactions, while its influence can be neglected for short-duration interactions. However, our simulations suggest that the importance of the gas influence depends on the dynamic details of the sBH close encounter process. With a high relative speed\,($\gg c_s$) during the encounter, the dense gas between the two encountering sBHs will lag behind the motion of the sBHs after the collision. The gravitational force exerted by the lagged gas on the two sBHs causes energy loss in the three-body system, thereby influencing the end states.

The end states of our simulation results are shown in the left panel of Figure\,\ref{fig:end_state}, of which mostly are in a BBH + single sBH state. Compared to \(\Sigma_0 = 0\), the presence of gas alters the end states of the BSI. For the BSI process with a fixed surface density $\Sigma_0$, we count the number of events with the end state changed by the presence of the dense gas $\Delta N_{\Sigma_0}$ compared to the gas-absent case $N_{\Sigma_0=0}$. Figure\,\ref{fig:end_state} shows an increasing ratio $\Delta N_{\Sigma_0}/N_{\Sigma_0=0}=(16/60,\,16/60,\,17/60,\,26/60,\,37/60)$ for $\Sigma_0=(1/5,\,1/\sqrt{5},\, 1,\,\sqrt{5},\, 5)\,\Sigma_\star$. Essentially, this is because both the duration of the BSI process and the number of intermittent encounters during BSI increase with the disk gas density. For example, the number of intermittent encounters is (4, 4, 8, 9, 14, 20) for $\Sigma_0=(0,\,1/5,\,1/\sqrt{5},\, 1,\,\sqrt{5},\, 5)\,\Sigma_\star$, respectively, for $b = 2.005\,R_{\rm H}$. In addition, the probability of the end state being an ST state gradually increases with higher gas density. This can be explained by the fact that higher gas density leads to higher energy dissipation in the three-body system.
 
Additionally, we show the orbital angular momentum directions of the end-state BBH in the right panel of Figure\,\ref{fig:end_state}, which are generally not significantly changed by the presence of the dense gas.  The percentage of end-state BBHs that are retrograde is approximately 20\% to 38\%, depending on the gas density. For the retrograde BBHs in AGN disk, they may experience runaway orbital eccentricity growth, resulting in a quick merger\,\citep[see][]{Calcino2024ApJ}.

\subsection{Orbital Parameters of the End-state BBHs}
Using the equation\,\eqref{eq:energy_binary}, we can calculate the semi-major axis $a_{\rm b}$ and eccentricity $e_{\rm b}$ of the end-state BBHs. Note that in this work, we do not simulate the subsequent evolution of end-state BBHs in AGN disk for two main reasons: (a) the pericenters of end-state BBHs could be extremely small, which require very high resolution to model accurately; (b) after the BSI, most residue BBHs move out of the shearing box. Therefore, we focus exclusively on BBHs at the end-state moment, defined as the time $T_2$ when the distance between two sBHs exceeds \( R_{\rm H} \) and remains so until the end of our simulation\,(\(3\,T_0\)). The subsequent evolution of these high-eccentricity BBHs can be anticipated based on previous studies\,\citep[e.g.,][]{Rowan2023MNRAS,Calcino2024ApJ}. For example, for prograde BBHs, eccentricity tends to dampen over time gradually. In contrast, retrograde BBHs may experience runaway eccentricity growth, leading to a rapid merger. Retrograde BBHs contract their orbits 3–4 times faster than prograde BBHs.

In Figure\,\ref{fig:af_ef}, we show the distribution of the $a_{\rm b}$ and $1-e_{\rm b}$ in parameter spaces. As the gas density increases, the semi-major axes of BBHs tend to be more compact due to the energy dissipation of the three-body system in a gas-rich environment. Since the influence of gas on the angular momentum of three-body system during the BSI process can be neglected, its effect on the eccentricity distribution is minor.

For a coplanar three-body system in a gas-free environment, the cumulative distributions\,(CDFs) of the end-state-BBH semi-major axis $a_{\rm b}$ and eccentricity $e_{\rm b}$ can be analytically given by\,\citep[e.g.,][]{Monaghan1976MNRAS,Valtonen2006tbp,Stone2019Natur}:
\begin{equation} \label{eq:CDF}
    \begin{split}
      &p(a<a_{\rm b}) = \int_0^{a_{\rm b}} p(a_{\rm b}) \,{\rm d}\, a_{\rm b} = a_{\rm max}^{-2} a_{\rm b}^2, \\
        & p(1-e<1-e_{\rm b}) = \int_0^{1-e_{\rm b}} p(1-e_{\rm b}) \,{\rm d}\, (1-e_{\rm b})\\
        &\qquad\qquad\quad\quad\quad = \sqrt{1-e_{\rm b}^2},
    \end{split}
\end{equation}
where $a_{\rm max}\sim a_{\rm b0}$ is the maximum semi-major axis of the end-state BBH. 
In the first panel of Figure\,\ref{fig:aeT_dis}, we compare the CDFs of the end-state BBHs at various densities with those predicted by analytical theory. The solid lines show the CDFs of end-state BBHs under various gas densities, while the black dashed lines in the first two panels represent the theoretically predicted distributions of the semi-major axis and eccentricity, which is given by equation\,\eqref{eq:CDF}. When the gas is absent or the density is low, the CDFs of the semi-major axis and the eccentricity of end-state BBHs are consistent with the analytical results. The minor deviations observed in our \(\Sigma_0 = 0\) simulations stem from three factors: (a) the analytical solution neglects the influence of the central SMBH, (b) insufficiently compact initial BBH separation\,($R_{\rm H}/5$) results in the three-body system retaining a dependence on initial conditions, preventing it from entering a fully chaotic state\,\citep[see][]{Fabj2024arXiv}, and (c) the limited sample size of 60 simulations for each disk density may introduce statistical fluctuations.
Figure\,\ref{fig:aeT_dis} suggests that 
the presence of gas significantly shifts the distribution of $a_{\rm b}$ towards a smaller semi-major axis, while the effect on the distribution of eccentricity is insignificant. 

Typically, the end-state-BBH pericenter $r_{\rm p}$ is used to substitute for the closest distance between sBHs during the BSI process\,\citep[see][ for details]{Valtonen2006tbp}. The timescale distribution of the end-state BBH merger\,($T_{\rm GW}$) can be estimated using the CDF of the $r_p$.  Therefore, in the third panel of Figure\,\ref{fig:aeT_dis}, we present the CDFs of $r_{\rm p}$ at different gas densities. We find that the presence of gas causes the CDF of the pericenter to shift toward smaller pericenter values. Meanwhile, we calculate the GW merger timescale of the end-state BBHs, which can be approximated in the high eccentricity limit by using
\begin{equation}
    T_{\rm GW} = \frac{2^{7/2} 5c^5}{512G^3}  m^{-3} a_{\rm b}^4 (1-e_{\rm b})^{7/2}.
\end{equation}
The CDF of the GW merger timescale is shown in the fourth panel of Figure\,\ref{fig:aeT_dis}. We find that the presence of gas significantly shortens the GW merger timescale of the end-state BBHs.

\section{conclusions AND discussions} \label{sec:discussions}
In this work, we investigate the effect of the dense gas on the BSI process in the AGN disk by coupling the hydrodynamical simulation and the N-body simulation. A total of 360 simulations are conducted, covering a range of gas surface densities $\Sigma_0$ and impact parameters $b$. Our simulations demonstrate the dynamical details of the evolution of the three-body system under the influence of dense gas.  The energy and angular momentum of the three-body system during the BSI process are quantitatively discussed for a fiducial model. Furthermore, we compare the orbital parameter distributions of end-state residue BBHs at different densities, statistically examining the impact of gas on the distribution of end-state BBHs. Our main conclusions can be summarized as follows:

(1) Compared to the gas-free environment, the gas environment can lead to a longer BSI duration and a higher number of close encounters. For the encounter process with sufficiently large relative sBH speed\,($\gg c_s$), the accumulated gas from the collision of circum-single disks may gravitationally drag the sBHs flying apart, which creates a significant energy dissipation from the three-body system to the dense gas. 

(2) The end states of the BSI process can be changed by the dense gas environment. The proportion of the end states differing from those in the absence of gas\,(\(\Sigma_0 = 0\)) ranges from 16/60 to 37/60 across the five different gas density levels.

(3) The semi-major axis of end-state BBHs tends to be more compact due to energy dissipation from the gas in the three-body system, while the eccentricity distribution remains nearly unaffected. In this way, the presence of gas shortens the GW merger timescale for end-state BBHs.

The following limitations of our model shounld be discussed:

(1) Due to the low proportion of direct mergers during the BSI process\,($\sim 4\%$) and the limited number of simulations (only 60 simulations for each density), we are unable to compare the rate of direct three-body mergers during the BSI process across different gas densities. This is also why we do not include GW radiation in the N-body simulations.
Instead, following previous analytical studies\,\citep[e.g.,][]{Valtonen2006tbp}, we can infer the impact of gas on the three-body merger rate during the BSI process from the number of encounters during BSI and the distribution of the end-state-BBH pericenter $r_{\rm p}$. As the gas density increases, the CDF of the pericenter shifts toward smaller values, which could indicate a higher merger rate. Meanwhile, the increase in the number of encounters also suggests that the rate of direct three-body mergers will be enhanced due to the presence of gas.
In future work, we will directly compare the impact of gas on the direct three-body merger rate by spanning a larger parameter space\,(Wang et al. in preparation). 

(2) In this work, we primarily present results for the case of a prograde initial BBH. Moreover, we have also explored scenarios with retrograde initial BBHs for specific parameter sets. We find that for the chaotic BSI processes, there is no fundamental difference between prograde and retrograde initial BBHs, which is consistent with the results in \cite{Fabj2024arXiv}.

(3) The three-dimensional\,(3D) hydrodynamical simulations, yet to be performed, may have the following impact on our results. Firstly, we believe that 3D simulations would not qualitatively alter the main conclusions drawn from 2D simulations. The fundamental mechanism of energy loss due to gas accumulation during sBH encounters remains qualitatively similar, only the high-density regions generated by collisions of CSDs transition from linear regions in 2D simulations to planar regions in 3D simulations. Secondly, in 3D simulations, if sBHs are no longer confined to the same plane, the eccentricity distribution of temporary BBHs formed during the BSI process shifts from a superthermal distribution (\( p(e_{\rm b}) = e_{\rm b}/\sqrt{1 - e_{\rm b}^2} \)) toward a thermal distribution (\( p(e_{\rm b}) = 2e_{\rm b} \)), significantly reducing the three-body merger rate\,\citep[see Figure 11 of][]{Fabj2024arXiv}.

(4) Up to now, the accretion of sBHs in an AGN disk remains poorly understood, which deserves further study in the future. Previous studies have treated gas accretion in different accretion methodologies and rate\,\citep[e.g.,][]{LRX2022MNRAS,Dempsey2022ApJ,Rowan2023MNRAS,Whitehead2024MNRAS}. In this work, we neglect the changes in mass and momentum of sBHs due to gas accretion for the following reasons: (a) Although it is challenging to determine the true accretion rate of sBHs limited by the resolution of the HD simulation and the complexity of feedback mechanisms, we expect that the mass of sBHs will not change significantly over the short timescale\,(\(\sim \text{year}\)) of the BSI process\,(the mass change of sBHs during our simulations $\Delta m / 50M_\odot  < 1\%$). (b)  Previous simulation works show that the gas-accretion-induced mass and momentum change is significantly smaller than the gravitational effects of the gas near the sBHs\,\citep[e.g.,][]{Calcino2024ApJ,LYP2024ApJ}. 

After completing this work, we noted a recently independent work on similar topics by \cite{Rowan2025arXiv}. They assumed three non-accreting sBHs embedding in the AGN disk, resulting in stronger energy dissipation during the BSI process. Readers are encouraged to compare our findings with those in \cite{Rowan2025arXiv}.

\begin{acknowledgments}
M.W. thanks Xiao Fan, and Di Wang for helpful discussions. M.W. and Q.W. gratefully acknowledge support from the National Natural Science Foundation of China (grants 123B2041 and 12233007), the National Key Research and Development Program of China (No. 2023YFC2206702) and the National SKA Program of China (2022SKA0120101). Y.M. is supported by the university start-up fund provided by Huazhong University of Science and Technology. Y.P.L. is supported in part by the Natural Science Foundation of China (grants 12373070, and 12192223), the Natural Science Foundation of Shanghai (grant NO. 23ZR1473700). The scope of H.L.'s research presented in this article is on the underlying physics and interpretation of the results and was supported by the Laboratory Directed Research and Development program of Los Alamos National Laboratory under project number 20220087DR. M.W. acknowledge Beijing PARATERA Tech CO., Ltd. for providing HPC resources that have contributed to the research results reported within this paper.
\end{acknowledgments}

\bibliography{ref.bib}{}

\begin{thebibliography}{}
\expandafter\ifx\csname natexlab\endcsname\relax\def\natexlab#1{#1}\fi
\providecommand{\url}[1]{\href{#1}{#1}}
\providecommand{\dodoi}[1]{doi:~\href{http://doi.org/#1}{\nolinkurl{#1}}}
\providecommand{\doeprint}[1]{\href{http://ascl.net/#1}{\nolinkurl{http://ascl.net/#1}}}
\providecommand{\doarXiv}[1]{\href{https://arxiv.org/abs/#1}{\nolinkurl{https://arxiv.org/abs/#1}}}

\bibitem[{{Abbott} {et~al.}(2020{\natexlab{a}}){Abbott}, {Abbott}, {Abraham},
  {Acernese}, {Ackley}, {Adams}, {Adhikari}, {Adya}, {Affeldt}, {Agathos},
  {Agatsuma}, {Aggarwal}, {Aguiar}, {Aich}, {Aiello}, {Ain}, {Ajith}, {Akcay},
  {Allen}, {Allocca}, {Altin}, {Amato}, {Anand}, {Ananyeva}, {Anderson},
  {Anderson}, {Angelova}, {Ansoldi}, {Antier}, {Appert}, {Arai}, {Araya},
  {Areeda}, {Ar{\`e}ne}, {Arnaud}, {Aronson}, {Arun}, {Asali}, {Ascenzi},
  {Ashton}, {Aston}, {Astone}, {Aubin}, {Aufmuth}, {AultONeal}, {Austin},
  {Avendano}, {Babak}, {Bacon}, {Badaracco}, {Bader}, {Bae}, {Baer}, {Baird},
  {Baldaccini}, {Ballardin}, {Ballmer}, {Bals}, {Balsamo}, {Baltus},
  {Banagiri}, {Bankar}, {Bankar}, {Barayoga}, {Barbieri}, {Barish}, {Barker},
  {Barkett}, {Barneo}, {Barone}, {Barr}, {Barsotti}, {Barsuglia}, {Barta},
  {Bartlett}, {Bartos}, {Bassiri}, {Basti}, {Bawaj}, {Bayley}, {Bazzan},
  {B{\'e}csy}, {Bejger}, {Belahcene}, {Bell}, {Beniwal}, {Benjamin}, {Bentley},
  {Bergamin}, {Berger}, {Bergmann}, {Bernuzzi}, {Berry}, {Bersanetti},
  {Bertolini}, {Betzwieser}, {Bhandare}, {Bhandari}, {Bidler}, {Biggs},
  {Bilenko}, {Billingsley}, {Birney}, {Birnholtz}, {Biscans}, {Bischi},
  {Biscoveanu}, {Bisht}, {Bissenbayeva}, {Bitossi}, {Bizouard}, {Blackburn},
  {Blackman}, {Blair}, {Blair}, {Blair}, {Bobba}, {Bode}, {Boer}, {Boetzel},
  {Bogaert}, {Bondu}, {Bonilla}, {Bonnand}, {Booker}, {Boom}, {Bork}, {Boschi},
  {Bose}, {Bossilkov}, {Bosveld}, {Bouffanais}, {Bozzi}, {Bradaschia}, {Brady},
  {Bramley}, {Branchesi}, {Brau}, {Breschi}, {Briant}, {Briggs}, {Brighenti},
  {Brillet}, {Brinkmann}, {Brockill}, {Brooks}, {Brooks}, {Brown}, {Brunett},
  {Bruno}, {Bruntz}, {Buikema}, {Bulik}, {Bulten}, {Buonanno}, {Buscicchio},
  {Buskulic}, {Byer}, {Cabero}, {Cadonati}, {Cagnoli}, {Cahillane},
  {Calder{\'o}n Bustillo}, {Callaghan}, {Callister}, {Calloni}, {Camp},
  {Canepa}, {Cannon}, {Cao}, {Cao}, {Carapella}, {Carbognani}, {Caride},
  {Carney}, {Carullo}, {Casanueva Diaz}, {Casentini}, {Casta{\~n}eda},
  {Caudill}, {Cavagli{\`a}}, {Cavalier}, {Cavalieri}, {Cella},
  {Cerd{\'a}-Dur{\'a}n}, {Cesarini}, {Chaibi}, {Chakravarti}, {Chan}, {Chan},
  {Chandra}, {Chao}, {Charlton}, {Chase}, {Chassande-Mottin}, {Chatterjee},
  {Chaturvedi}, {Chatziioannou}, {Chen}, {Chen}, {Chen}, {Cheng}, {Cheong},
  {Chia}, {Chiadini}, {Chierici}, {Chincarini}, {Chiummo}, {Cho}, {Cho}, {Cho},
  {Christensen}, {Chu}, {Chua}, {Chung}, {Chung}, {Ciani}, {Ciecielag},
  {Cie{\'s}lar}, {Ciobanu}, {Ciolfi}, {Cipriano}, {Cirone}, {Clara}, {Clark},
  {Clearwater}, {Clesse}, {Cleva}, {Coccia}, {Cohadon}, {Cohen}, {Colleoni},
  {Collette}, {Collins}, {Colpi}, {Constancio}, {Conti}, {Cooper}, {Corban},
  {Corbitt}, {Cordero-Carri{\'o}n}, {Corezzi}, {Corley}, {Cornish}, {Corre},
  {Corsi}, {Cortese}, {Costa}, {Cotesta}, {Coughlin}, {Coughlin}, {Coulon},
  {Countryman}, {Couvares}, {Covas}, {Coward}, {Cowart}, {Coyne}, {Coyne},
  {Creighton}, {Creighton}, {Cripe}, {Croquette}, {Crowder}, {Cudell},
  {Cullen}, {Cumming}, {Cummings}, {Cunningham}, {Cuoco}, {Curylo}, {Canton},
  {D{\'a}lya}, {Dana}, {Daneshgaran-Bajastani}, {D'Angelo}, {Danilishin},
  {D'Antonio}, {Danzmann}, {Darsow-Fromm}, {Dasgupta}, {Datrier}, {Dattilo},
  {Dave}, {Davier}, {Davies}, {Davis}, {Daw}, {DeBra}, {Deenadayalan},
  {Degallaix}, {De Laurentis}, {Del{\'e}glise}, {Delfavero}, {De Lillo}, {Del
  Pozzo}, {DeMarchi}, {D'Emilio}, {Demos}, {Dent}, {De Pietri}, {De Rosa}, {De
  Rossi}, {DeSalvo}, {de Varona}, {Dhurandhar}, {D{\'\i}az}, {Diaz-Ortiz},
  {Dietrich}, {Di Fiore}, {Di Fronzo}, {Di Giorgio}, {Di Giovanni}, {Di
  Giovanni}, {Di Girolamo}, {Di Lieto}, {Ding}, {Di Pace}, {Di Palma}, {Di
  Renzo}, {Divakarla}, {Dmitriev}, {Doctor}, {Donovan}, {Dooley}, {Doravari},
  {Dorrington}, {Downes}, {Drago}, {Driggers}, {Du}, {Ducoin}, {Dupej},
  {Durante}, {D'Urso}, {Dwyer}, {Easter}, {Eddolls}, {Edelman}, {Edo}, {Edy},
  {Effler}, {Ehrens}, {Eichholz}, {Eikenberry}, {Eisenmann}, {Eisenstein},
  {Ejlli}, {Errico}, {Essick}, {Estelles}, {Estevez}, {Etienne}, {Etzel},
  {Evans}, {Evans}, {Ewing}, {Fafone}, {Fairhurst}, {Fan}, {Farinon}, {Farr},
  {Farr}, {Fauchon-Jones}, {Favata}, {Fays}, {Fazio}, {Feicht}, {Fejer},
  {Feng}, {Fenyvesi}, {Ferguson}, {Fernandez-Galiana}, {Ferrante}, {Ferreira},
  {Ferreira}, {Fidecaro}, {Fiori}, {Fiorucci}, {Fishbach}, {Fisher},
  {Fittipaldi}, {Fitz-Axen}, {Fiumara}, {Flaminio}, {Floden}, {Flynn}, {Fong},
  {Font}, {Forsyth}, {Fournier}, {Frasca}, {Frasconi}, {Frei}, {Freise},
  {Frey}, {Frey}, {Fritschel}, {Frolov}, {Fronz{\`e}}, {Fulda}, {Fyffe},
  {Gabbard}, {Gadre}, {Gaebel}, {Gair}, {Galaudage}, {Ganapathy}, {Ganguly},
  {Gaonkar}, {Garc{\'\i}a-Quir{\'o}s}, {Garufi}, {Gateley}, {Gaudio},
  {Gayathri}, {Gemme}, {Genin}, {Gennai}, {George}, {George}, {Gergely},
  {Ghonge}, {Ghosh}, {Ghosh}, {Ghosh}, {Giacomazzo}, {Giaime}, {Giardina},
  {Gibson}, {Gier}, {Gill}, {Glanzer}, {Gniesmer}, {Godwin}, {Goetz}, {Goetz},
  {Gohlke}, {Goncharov}, {Gonz{\'a}lez}, {Gopakumar}, {Gossan}, {Gosselin},
  {Gouaty}, {Grace}, {Grado}, {Granata}, {Grant}, {Gras}, {Grassia}, {Gray},
  {Gray}, {Greco}, {Green}, {Green}, {Gretarsson}, {Griggs}, {Grignani},
  {Grimaldi}, {Grimm}, {Grote}, {Grunewald}, {Gruning}, {Guidi}, {Guimaraes},
  {Guix{\'e}}, {Gulati}, {Guo}, {Gupta}, {Gupta}, {Gupta}, {Gustafson},
  {Gustafson}, {Haegel}, {Halim}, {Hall}, {Hamilton}, {Hammond}, {Haney},
  {Hanke}, {Hanks}, {Hanna}, {Hannam}, {Hannuksela}, {Hansen}, {Hanson},
  {Harder}, {Hardwick}, {Haris}, {Harms}, {Harry}, {Harry}, {Hasskew},
  {Haster}, {Haughian}, {Hayes}, {Healy}, {Heidmann}, {Heintze}, {Heinze},
  {Heitmann}, {Hellman}, {Hello}, {Hemming}, {Hendry}, {Heng}, {Hennes},
  {Hennig}, {Heurs}, {Hild}, {Hinderer}, {Hoback}, {Hochheim}, {Hofgard},
  {Hofman}, {Holgado}, {Holland}, {Holt}, {Holz}, {Hopkins}, {Horst}, {Hough},
  {Howell}, {Hoy}, {Huang}, {H{\"u}bner}, {Huerta}, {Huet}, {Hughey}, {Hui},
  {Husa}, {Huttner}, {Huxford}, {Huynh-Dinh}, {Idzkowski}, {Iess}, {Inchauspe},
  {Ingram}, {Intini}, {Isac}, {Isi}, {Iyer}, {Jacqmin}, {Jadhav}, {Jadhav},
  {James}, {Jani}, {Janthalur}, {Jaranowski}, {Jariwala}, {Jaume}, {Jenkins},
  {Jiang}, {Johns}, {Johnson-McDaniel}, {Jones}, {Jones}, {Jones}, {Jones},
  {Jones}, {Jonker}, {Ju}, {Junker}, {Kalaghatgi}, {Kalogera}, {Kamai},
  {Kandhasamy}, {Kang}, {Kanner}, {Kapadia}, {Karki}, {Kashyap}, {Kasprzack},
  {Kastaun}, {Katsanevas}, {Katsavounidis}, {Katzman}, {Kaufer}, {Kawabe},
  {K{\'e}f{\'e}lian}, {Keitel}, {Keivani}, {Kennedy}, {Key}, {Khadka},
  {Khalili}, {Khan}, {Khan}, {Khan}, {Khazanov}, {Khetan}, {Khursheed},
  {Kijbunchoo}, {Kim}, {Kim}, {Kim}, {Kim}, {Kim}, {Kim}, {Kim}, {Kimball},
  {King}, {Kinley-Hanlon}, {Kirchhoff}, {Kissel}, {Kleybolte}, {Klimenko},
  {Knowles}, {Knyazev}, {Koch}, {Koehlenbeck}, {Koekoek}, {Koley},
  {Kondrashov}, {Kontos}, {Koper}, {Korobko}, {Korth}, {Kovalam}, {Kozak},
  {Kringel}, {Krishnendu}, {Kr{\'o}lak}, {Krupinski}, {Kuehn}, {Kumar},
  {Kumar}, {Kumar}, {Kumar}, {Kumar}, {Kuo}, {Kutynia}, {Lackey}, {Laghi},
  {Lalande}, {Lam}, {Lamberts}, {Landry}, {Lane}, {Lang}, {Lange}, {Lantz},
  {Lanza}, {La Rosa}, {Lartaux-Vollard}, {Lasky}, {Laxen}, {Lazzarini},
  {Lazzaro}, {Leaci}, {Leavey}, {Lecoeuche}, {Lee}, {Lee}, {Lee}, {Lee}, {Lee},
  {Lehmann}, {Leroy}, {Letendre}, {Levin}, {Li}, {Li}, {li}, {Li}, {Li},
  {Linde}, {Linker}, {Linley}, {Littenberg}, {Liu}, {Liu},
  {Llorens-Monteagudo}, {Lo}, {Lockwood}, {London}, {Longo}, {Lorenzini},
  {Loriette}, {Lormand}, {Losurdo}, {Lough}, {Lousto}, {Lovelace}, {L{\"u}ck},
  {Lumaca}, {Lundgren}, {Ma}, {Macas}, {Macfoy}, {MacInnis}, {Macleod},
  {MacMillan}, {Macquet}, {Maga{\~n}a Hernandez}, {Maga{\~n}a-Sandoval},
  {Magee}, {Majorana}, {Maksimovic}, {Malik}, {Man}, {Mandic}, {Mangano},
  {Mansell}, {Manske}, {Mantovani}, {Mapelli}, {Marchesoni}, {Marion},
  {M{\'a}rka}, {M{\'a}rka}, {Markakis}, {Markosyan}, {Markowitz}, {Maros},
  {Marquina}, {Marsat}, {Martelli}, {Martin}, {Martin}, {Martinez}, {Martynov},
  {Masalehdan}, {Mason}, {Massera}, {Masserot}, {Massinger}, {Masso-Reid},
  {Mastrogiovanni}, {Matas}, {Matichard}, {Mavalvala}, {Maynard}, {McCann},
  {McCarthy}, {McClelland}, {McCormick}, {McCuller}, {McGuire}, {McIsaac},
  {McIver}, {McManus}, {McRae}, {McWilliams}, {Meacher}, {Meadors}, {Mehmet},
  {Mehta}, {Mejuto Villa}, {Melatos}, {Mendell}, {Mercer}, {Mereni}, {Merfeld},
  {Merilh}, {Merritt}, {Merzougui}, {Meshkov}, {Messenger}, {Messick},
  {Metzdorff}, {Meyers}, {Meylahn}, {Mhaske}, {Miani}, {Miao}, {Michaloliakos},
  {Michel}, {Middleton}, {Milano}, {Miller}, {Millhouse}, {Mills}, {Milotti},
  {Milovich-Goff}, {Minazzoli}, {Minenkov}, {Mishkin}, {Mishra}, {Mistry},
  {Mitra}, {Mitrofanov}, {Mitselmakher}, {Mittleman}, {Mo}, {Mogushi},
  {Mohapatra}, {Mohite}, {Molina-Ruiz}, {Mondin}, {Montani}, {Moore}, {Moraru},
  {Morawski}, {Moreno}, {Morisaki}, {Mours}, {Mow-Lowry}, {Mozzon},
  {Muciaccia}, {Mukherjee}, {Mukherjee}, {Mukherjee}, {Mukherjee}, {Mukund},
  {Mullavey}, {Munch}, {Mu{\~n}iz}, {Murray}, {Nagar}, {Nardecchia},
  {Naticchioni}, {Nayak}, {Neil}, {Neilson}, {Nelemans}, {Nelson}, {Nery},
  {Neunzert}, {Ng}, {Ng}, {Nguyen}, {Nguyen}, {Nichols}, {Nichols}, {Nissanke},
  {Nitz}, {Nocera}, {Noh}, {North}, {Nothard}, {Nuttall}, {Oberling},
  {O'Brien}, {Oganesyan}, {Ogin}, {Oh}, {Oh}, {Ohme}, {Ohta}, {Okada},
  {Oliver}, {Olivetto}, {Oppermann}, {Oram}, {O'Reilly}, {Ormiston}, {Ortega},
  {O'Shaughnessy}, {Ossokine}, {Osthelder}, {Ottaway}, {Overmier}, {Owen},
  {Pace}, {Pagano}, {Page}, {Pagliaroli}, {Pai}, {Pai}, {Palamos}, {Palashov},
  {Palomba}, {Pan}, {Panda}, {Pang}, {Pankow}, {Pannarale}, {Pant}, {Paoletti},
  {Paoli}, {Parida}, {Parker}, {Pascucci}, {Pasqualetti}, {Passaquieti},
  {Passuello}, {Patricelli}, {Payne}, {Pearlstone}, {Pechsiri}, {Pedersen},
  {Pedraza}, {Pele}, {Penn}, {Perego}, {Perez}, {P{\'e}rigois}, {Perreca},
  {Perri{\`e}s}, {Petermann}, {Pfeiffer}, {Phelps}, {Phukon}, {Piccinni},
  {Pichot}, {Piendibene}, {Piergiovanni}, {Pierro}, {Pillant}, {Pinard},
  {Pinto}, {Piotrzkowski}, {Pirello}, {Pitkin}, {Plastino}, {Poggiani}, {Pong},
  {Ponrathnam}, {Popolizio}, {Porter}, {Powell}, {Prajapati}, {Prasai},
  {Prasanna}, {Pratten}, {Prestegard}, {Principe}, {Prodi}, {Prokhorov},
  {Punturo}, {Puppo}, {P{\"u}rrer}, {Qi}, {Quetschke}, {Quinonez}, {Raab},
  {Raaijmakers}, {Radkins}, {Radulesco}, {Raffai}, {Rafferty}, {Raja}, {Rajan},
  {Rajbhandari}, {Rakhmanov}, {Ramirez}, {Ramos-Buades}, {Rana}, {Rao},
  {Rapagnani}, {Raymond}, {Razzano}, {Read}, {Regimbau}, {Rei}, {Reid},
  {Reitze}, {Rettegno}, {Ricci}, {Richardson}, {Richardson}, {Ricker},
  {Riemenschneider}, {Riles}, {Rizzo}, {Robertson}, {Robinet}, {Rocchi},
  {Rodriguez-Soto}, {Rolland}, {Rollins}, {Roma}, {Romanelli}, {Romano},
  {Romel}, {Romero-Shaw}, {Romie}, {Rose}, {Rose}, {Rose}, {Rosi{\'n}ska},
  {Rosofsky}, {Ross}, {Rowan}, {Rowlinson}, {Roy}, {Roy}, {Roy}, {Ruggi},
  {Rutins}, {Ryan}, {Sachdev}, {Sadecki}, {Sakellariadou}, {Salafia},
  {Salconi}, {Saleem}, {Salemi}, {Samajdar}, {Sanchez}, {Sanchez},
  {Sanchis-Gual}, {Sanders}, {Santiago}, {Santos}, {Sarin}, {Sassolas},
  {Sathyaprakash}, {Sauter}, {Savage}, {Savant}, {Sawant}, {Sayah}, {Schaetzl},
  {Schale}, {Scheel}, {Scheuer}, {Schmidt}, {Schnabel}, {Schofield},
  {Sch{\"o}nbeck}, {Schreiber}, {Schulte}, {Schutz}, {Schwarm}, {Schwartz},
  {Scott}, {Scott}, {Seidel}, {Sellers}, {Sengupta}, {Sennett}, {Sentenac},
  {Sequino}, {Sergeev}, {Setyawati}, {Shaddock}, {Shaffer}, {Sharifi},
  {Shahriar}, {Sharma}, {Sharma}, {Shawhan}, {Shen}, {Shikauchi}, {Shink},
  {Shoemaker}, {Shoemaker}, {Shukla}, {ShyamSundar}, {Siellez}, {Sieniawska},
  {Sigg}, {Singer}, {Singh}, {Singh}, {Singha}, {Singhal}, {Sintes}, {Sipala},
  {Skliris}, {Slagmolen}, {Slaven-Blair}, {Smetana}, {Smith}, {Smith},
  {Somala}, {Son}, {Soni}, {Sorazu}, {Sordini}, {Sorrentino}, {Souradeep},
  {Sowell}, {Spencer}, {Spera}, {Srivastava}, {Srivastava}, {Staats},
  {Stachie}, {Standke}, {Steer}, {Steinke}, {Steinlechner}, {Steinlechner},
  {Steinmeyer}, {Stevenson}, {Stocks}, {Stops}, {Stover}, {Strain}, {Stratta},
  {Strunk}, {Sturani}, {Stuver}, {Sudhagar}, {Sudhir}, {Summerscales}, {Sun},
  {Sunil}, {Sur}, {Suresh}, {Sutton}, {Swinkels}, {Szczepa{\'n}czyk}, {Tacca},
  {Tait}, {Talbot}, {Tanasijczuk}, {Tanner}, {Tao}, {T{\'a}pai}, {Tapia},
  {Tapia San Martin}, {Tasson}, {Taylor}, {Tenorio}, {Terkowski},
  {Thirugnanasambandam}, {Thomas}, {Thomas}, {Thompson}, {Thondapu}, {Thorne},
  {Thrane}, {Tinsman}, {Saravanan}, {Tiwari}, {Tiwari}, {Tiwari}, {Toland},
  {Tonelli}, {Tornasi}, {Torres-Forn{\'e}}, {Torrie}, {Tosta e Melo},
  {T{\"o}yr{\"a}}, {Travasso}, {Traylor}, {Tringali}, {Tripathee}, {Trovato},
  {Trudeau}, {Tsang}, {Tse}, {Tso}, {Tsukada}, {Tsuna}, {Tsutsui}, {Turconi},
  {Ubhi}, {Udall}, {Ueno}, {Ugolini}, {Unnikrishnan}, {Urban}, {Usman},
  {Utina}, {Vahlbruch}, {Vajente}, {Valdes}, {Valentini}, {van Bakel}, {van
  Beuzekom}, {van den Brand}, {Van Den Broeck}, {Vander-Hyde}, {van der
  Schaaf}, {Van Heijningen}, {van Veggel}, {Vardaro}, {Varma}, {Vass},
  {Vas{\'u}th}, {Vecchio}, {Vedovato}, {Veitch}, {Veitch}, {Venkateswara},
  {Venugopalan}, {Verkindt}, {Veske}, {Vetrano}, {Vicer{\'e}}, {Viets},
  {Vinciguerra}, {Vine}, {Vinet}, {Vitale}, {Vivanco}, {Vo}, {Vocca},
  {Vorvick}, {Vyatchanin}, {Wade}, {Wade}, {Wade}, {Walet}, {Walker},
  {Wallace}, {Wallace}, {Walsh}, {Wang}, {Wang}, {Wang}, {Ward}, {Warden},
  {Warner}, {Was}, {Watchi}, {Weaver}, {Wei}, {Weinert}, {Weinstein}, {Weiss},
  {Wellmann}, {Wen}, {We{\ss}els}, {Westhouse}, {Wette}, {Whelan}, {Whiting},
  {Whittle}, {Wilken}, {Williams}, {Willis}, {Willke}, {Winkler}, {Wipf},
  {Wittel}, {Woan}, {Woehler}, {Wofford}, {Wong}, {Wright}, {Wu}, {Wysocki},
  {Xiao}, {Yamamoto}, {Yang}, {Yang}, {Yang}, {Yap}, {Yazback}, {Yeeles}, {Yu},
  {Yu}, {Yuen}, {Zadro{\.Z}ny}, {Zadro{\.Z}ny}, {Zanolin}, {Zelenova},
  {Zendri}, {Zevin}, {Zhang}, {Zhang}, {Zhang}, {Zhao}, {Zhao}, {Zhou}, {Zhou},
  {Zhu}, {Zimmerman}, {Zucker}, {Zweizig}, {LIGO Scientific Collaboration}, \&
  {Virgo Collaboration}}]{Abbott2020PhRvL}
{Abbott}, R., {Abbott}, T.~D., {Abraham}, S., {et~al.} 2020{\natexlab{a}},
  \prl, 125, 101102, \dodoi{10.1103/PhysRevLett.125.101102}

\bibitem[{{Abbott} {et~al.}(2020{\natexlab{b}}){Abbott}, {Abbott}, {Abraham},
  {Acernese}, {Ackley}, {Adams}, {Adhikari}, {Adya}, {Affeldt}, {Agathos},
  {Agatsuma}, {Aggarwal}, {Aguiar}, {Aich}, {Aiello}, {Ain}, {Ajith}, {Akcay},
  {Allen}, {Allocca}, {Altin}, {Amato}, {Anand}, {Ananyeva}, {Anderson},
  {Anderson}, {Angelova}, {Ansoldi}, {Antier}, {Appert}, {Arai}, {Araya},
  {Areeda}, {Ar{\`e}ne}, {Arnaud}, {Aronson}, {Arun}, {Asali}, {Ascenzi},
  {Ashton}, {Aston}, {Astone}, {Aubin}, {Aufmuth}, {AultONeal}, {Austin},
  {Avendano}, {Babak}, {Bacon}, {Badaracco}, {Bader}, {Bae}, {Baer}, {Baird},
  {Baldaccini}, {Ballardin}, {Ballmer}, {Bals}, {Balsamo}, {Baltus},
  {Banagiri}, {Bankar}, {Bankar}, {Barayoga}, {Barbieri}, {Barish}, {Barker},
  {Barkett}, {Barneo}, {Barone}, {Barr}, {Barsotti}, {Barsuglia}, {Barta},
  {Bartlett}, {Bartos}, {Bassiri}, {Basti}, {Bawaj}, {Bayley}, {Bazzan},
  {B{\'e}csy}, {Bejger}, {Belahcene}, {Bell}, {Beniwal}, {Benjamin}, {Bentley},
  {Bergamin}, {Berger}, {Bergmann}, {Bernuzzi}, {Berry}, {Bersanetti},
  {Bertolini}, {Betzwieser}, {Bhandare}, {Bhandari}, {Bidler}, {Biggs},
  {Bilenko}, {Billingsley}, {Birney}, {Birnholtz}, {Biscans}, {Bischi},
  {Biscoveanu}, {Bisht}, {Bissenbayeva}, {Bitossi}, {Bizouard}, {Blackburn},
  {Blackman}, {Blair}, {Blair}, {Blair}, {Bobba}, {Bode}, {Boer}, {Boetzel},
  {Bogaert}, {Bondu}, {Bonilla}, {Bonnand}, {Booker}, {Boom}, {Bork}, {Boschi},
  {Bose}, {Bossilkov}, {Bosveld}, {Bouffanais}, {Bozzi}, {Bradaschia}, {Brady},
  {Bramley}, {Branchesi}, {Brau}, {Breschi}, {Briant}, {Briggs}, {Brighenti},
  {Brillet}, {Brinkmann}, {Brockill}, {Brooks}, {Brooks}, {Brown}, {Brunett},
  {Bruno}, {Bruntz}, {Buikema}, {Bulik}, {Bulten}, {Buonanno}, {Buscicchio},
  {Buskulic}, {Byer}, {Cabero}, {Cadonati}, {Cagnoli}, {Cahillane}, {Bustillo},
  {Callaghan}, {Callister}, {Calloni}, {Camp}, {Canepa}, {Cannon}, {Cao},
  {Cao}, {Carapella}, {Carbognani}, {Caride}, {Carney}, {Carullo}, {Diaz},
  {Casentini}, {Casta{\~n}eda}, {Caudill}, {Cavagli{\`a}}, {Cavalier},
  {Cavalieri}, {Cella}, {Cerd{\'a}-Dur{\'a}n}, {Cesarini}, {Chaibi},
  {Chakravarti}, {Chan}, {Chan}, {Chao}, {Charlton}, {Chase},
  {Chassande-Mottin}, {Chatterjee}, {Chaturvedi}, {Chatziioannou}, {Chen},
  {Chen}, {Chen}, {Cheng}, {Cheong}, {Chia}, {Chiadini}, {Chierici},
  {Chincarini}, {Chiummo}, {Cho}, {Cho}, {Cho}, {Christensen}, {Chu}, {Chua},
  {Chung}, {Chung}, {Ciani}, {Ciecielag}, {Cie{\'s}lar}, {Ciobanu}, {Ciolfi},
  {Cipriano}, {Cirone}, {Clara}, {Clark}, {Clearwater}, {Clesse}, {Cleva},
  {Coccia}, {Cohadon}, {Cohen}, {Colleoni}, {Collette}, {Collins}, {Colpi},
  {Constancio}, {Conti}, {Cooper}, {Corban}, {Corbitt}, {Cordero-Carri{\'o}n},
  {Corezzi}, {Corley}, {Cornish}, {Corre}, {Corsi}, {Cortese}, {Costa},
  {Cotesta}, {Coughlin}, {Coughlin}, {Coulon}, {Countryman}, {Couvares},
  {Covas}, {Coward}, {Cowart}, {Coyne}, {Coyne}, {Creighton}, {Creighton},
  {Cripe}, {Croquette}, {Crowder}, {Cudell}, {Cullen}, {Cumming}, {Cummings},
  {Cunningham}, {Cuoco}, {Curylo}, {Canton}, {D{\'a}lya}, {Dana},
  {Daneshgaran-Bajastani}, {D'Angelo}, {Danilishin}, {D'Antonio}, {Danzmann},
  {Darsow-Fromm}, {Dasgupta}, {Datrier}, {Dattilo}, {Dave}, {Davier}, {Davies},
  {Davis}, {Daw}, {DeBra}, {Deenadayalan}, {Degallaix}, {De Laurentis},
  {Del{\'e}glise}, {Delfavero}, {De Lillo}, {Del Pozzo}, {DeMarchi},
  {D'Emilio}, {Demos}, {Dent}, {De Pietri}, {De Rosa}, {De Rossi}, {DeSalvo},
  {de Varona}, {Dhurandhar}, {D{\'\i}az}, {Diaz-Ortiz}, {Dietrich}, {Di Fiore},
  {Di Fronzo}, {Di Giorgio}, {Di Giovanni}, {Di Giovanni}, {Di Girolamo}, {Di
  Lieto}, {Ding}, {Di Pace}, {Di Palma}, {Di Renzo}, {Divakarla}, {Dmitriev},
  {Doctor}, {Donovan}, {Dooley}, {Doravari}, {Dorrington}, {Downes}, {Drago},
  {Driggers}, {Du}, {Ducoin}, {Dupej}, {Durante}, {D'Urso}, {Dwyer}, {Easter},
  {Eddolls}, {Edelman}, {Edo}, {Edy}, {Effler}, {Ehrens}, {Eichholz},
  {Eikenberry}, {Eisenmann}, {Eisenstein}, {Ejlli}, {Errico}, {Essick},
  {Estelles}, {Estevez}, {Etienne}, {Etzel}, {Evans}, {Evans}, {Ewing},
  {Fafone}, {Fairhurst}, {Fan}, {Farinon}, {Farr}, {Farr}, {Fauchon-Jones},
  {Favata}, {Fays}, {Fazio}, {Feicht}, {Fejer}, {Feng}, {Fenyvesi}, {Ferguson},
  {Fernandez-Galiana}, {Ferrante}, {Ferreira}, {Ferreira}, {Fidecaro}, {Fiori},
  {Fiorucci}, {Fishbach}, {Fisher}, {Fittipaldi}, {Fitz-Axen}, {Fiumara},
  {Flaminio}, {Floden}, {Flynn}, {Fong}, {Font}, {Forsyth}, {Fournier},
  {Frasca}, {Frasconi}, {Frei}, {Freise}, {Frey}, {Frey}, {Fritschel},
  {Frolov}, {Fronz{\`e}}, {Fulda}, {Fyffe}, {Gabbard}, {Gadre}, {Gaebel},
  {Gair}, {Galaudage}, {Ganapathy}, {Gaonkar}, {Garc{\'\i}a-Quir{\'o}s},
  {Garufi}, {Gateley}, {Gaudio}, {Gayathri}, {Gemme}, {Genin}, {Gennai},
  {George}, {George}, {Gergely}, {Ghonge}, {Ghosh}, {Ghosh}, {Ghosh},
  {Giacomazzo}, {Giaime}, {Giardina}, {Gibson}, {Gier}, {Gill}, {Glanzer},
  {Gniesmer}, {Godwin}, {Goetz}, {Goetz}, {Gohlke}, {Goncharov},
  {Gonz{\'a}lez}, {Gopakumar}, {Gossan}, {Gosselin}, {Gouaty}, {Grace},
  {Grado}, {Granata}, {Grant}, {Gras}, {Grassia}, {Gray}, {Gray}, {Greco},
  {Green}, {Green}, {Gretarsson}, {Griggs}, {Grignani}, {Grimaldi}, {Grimm},
  {Grote}, {Grunewald}, {Gruning}, {Guidi}, {Guimaraes}, {Guix{\'e}}, {Gulati},
  {Guo}, {Gupta}, {Gupta}, {Gupta}, {Gustafson}, {Gustafson}, {Haegel},
  {Halim}, {Hall}, {Hamilton}, {Hammond}, {Haney}, {Hanke}, {Hanks}, {Hanna},
  {Hannam}, {Hannuksela}, {Hansen}, {Hanson}, {Harder}, {Hardwick}, {Haris},
  {Harms}, {Harry}, {Harry}, {Hasskew}, {Haster}, {Haughian}, {Hayes}, {Healy},
  {Heidmann}, {Heintze}, {Heinze}, {Heitmann}, {Hellman}, {Hello}, {Hemming},
  {Hendry}, {Heng}, {Hennes}, {Hennig}, {Heurs}, {Hild}, {Hinderer}, {Hoback},
  {Hochheim}, {Hofgard}, {Hofman}, {Holgado}, {Holland}, {Holt}, {Holz},
  {Hopkins}, {Horst}, {Hough}, {Howell}, {Hoy}, {Huang}, {H{\"u}bner},
  {Huerta}, {Huet}, {Hughey}, {Hui}, {Husa}, {Huttner}, {Huxford},
  {Huynh-Dinh}, {Idzkowski}, {Iess}, {Inchauspe}, {Ingram}, {Intini}, {Isac},
  {Isi}, {Iyer}, {Jacqmin}, {Jadhav}, {Jadhav}, {James}, {Jani}, {Janthalur},
  {Jaranowski}, {Jariwala}, {Jaume}, {Jenkins}, {Jiang}, {Johns},
  {Johnson-McDaniel}, {Jones}, {Jones}, {Jones}, {Jones}, {Jones}, {Jonker},
  {Ju}, {Junker}, {Kalaghatgi}, {Kalogera}, {Kamai}, {Kandhasamy}, {Kang},
  {Kanner}, {Kapadia}, {Karki}, {Kashyap}, {Kasprzack}, {Kastaun},
  {Katsanevas}, {Katsavounidis}, {Katzman}, {Kaufer}, {Kawabe},
  {K{\'e}f{\'e}lian}, {Keitel}, {Keivani}, {Kennedy}, {Key}, {Khadka},
  {Khalili}, {Khan}, {Khan}, {Khan}, {Khazanov}, {Khetan}, {Khursheed},
  {Kijbunchoo}, {Kim}, {Kim}, {Kim}, {Kim}, {Kim}, {Kim}, {Kim}, {Kimball},
  {King}, {Kinley-Hanlon}, {Kirchhoff}, {Kissel}, {Kleybolte}, {Klimenko},
  {Knowles}, {Knyazev}, {Koch}, {Koehlenbeck}, {Koekoek}, {Koley},
  {Kondrashov}, {Kontos}, {Koper}, {Korobko}, {Korth}, {Kovalam}, {Kozak},
  {Kringel}, {Krishnendu}, {Kr{\'o}lak}, {Krupinski}, {Kuehn}, {Kumar},
  {Kumar}, {Kumar}, {Kumar}, {Kumar}, {Kuo}, {Kutynia}, {Lackey}, {Laghi},
  {Lalande}, {Lam}, {Lamberts}, {Landry}, {Lane}, {Lang}, {Lange}, {Lantz},
  {Lanza}, {La Rosa}, {Lartaux-Vollard}, {Lasky}, {Laxen}, {Lazzarini},
  {Lazzaro}, {Leaci}, {Leavey}, {Lecoeuche}, {Lee}, {Lee}, {Lee}, {Lee}, {Lee},
  {Lehmann}, {Leroy}, {Letendre}, {Levin}, {Li}, {Li}, {li}, {Li}, {Li},
  {Linde}, {Linker}, {Linley}, {Littenberg}, {Liu}, {Liu},
  {Llorens-Monteagudo}, {Lo}, {Lockwood}, {London}, {Longo}, {Lorenzini},
  {Loriette}, {Lormand}, {Losurdo}, {Lough}, {Lousto}, {Lovelace}, {L{\"u}ck},
  {Lumaca}, {Lundgren}, {Ma}, {Macas}, {Macfoy}, {MacInnis}, {Macleod},
  {MacMillan}, {Macquet}, {Hernandez}, {Maga{\~n}a-Sandoval}, {Magee},
  {Majorana}, {Maksimovic}, {Malik}, {Man}, {Mandic}, {Mangano}, {Mansell},
  {Manske}, {Mantovani}, {Mapelli}, {Marchesoni}, {Marion}, {M{\'a}rka},
  {M{\'a}rka}, {Markakis}, {Markosyan}, {Markowitz}, {Maros}, {Marquina},
  {Marsat}, {Martelli}, {Martin}, {Martin}, {Martinez}, {Martynov},
  {Masalehdan}, {Mason}, {Massera}, {Masserot}, {Massinger}, {Masso-Reid},
  {Mastrogiovanni}, {Matas}, {Matichard}, {Mavalvala}, {Maynard}, {McCann},
  {McCarthy}, {McClelland}, {McCormick}, {McCuller}, {McGuire}, {McIsaac},
  {McIver}, {McManus}, {McRae}, {McWilliams}, {Meacher}, {Meadors}, {Mehmet},
  {Mehta}, {Villa}, {Melatos}, {Mendell}, {Mercer}, {Mereni}, {Merfeld},
  {Merilh}, {Merritt}, {Merzougui}, {Meshkov}, {Messenger}, {Messick},
  {Metzdorff}, {Meyers}, {Meylahn}, {Mhaske}, {Miani}, {Miao}, {Michaloliakos},
  {Michel}, {Middleton}, {Milano}, {Miller}, {Millhouse}, {Mills}, {Milotti},
  {Milovich-Goff}, {Minazzoli}, {Minenkov}, {Mishkin}, {Mishra}, {Mistry},
  {Mitra}, {Mitrofanov}, {Mitselmakher}, {Mittleman}, {Mo}, {Mogushi},
  {Mohapatra}, {Mohite}, {Molina-Ruiz}, {Mondin}, {Montani}, {Moore}, {Moraru},
  {Morawski}, {Moreno}, {Morisaki}, {Mours}, {Mow-Lowry}, {Mozzon},
  {Muciaccia}, {Mukherjee}, {Mukherjee}, {Mukherjee}, {Mukherjee}, {Mukund},
  {Mullavey}, {Munch}, {Mu{\~n}iz}, {Murray}, {Nagar}, {Nardecchia},
  {Naticchioni}, {Nayak}, {Neil}, {Neilson}, {Nelemans}, {Nelson}, {Nery},
  {Neunzert}, {Ng}, {Ng}, {Nguyen}, {Nguyen}, {Nichols}, {Nichols}, {Nissanke},
  {Nocera}, {Noh}, {North}, {Nothard}, {Nuttall}, {Oberling}, {O'Brien},
  {Oganesyan}, {Ogin}, {Oh}, {Oh}, {Ohme}, {Ohta}, {Okada}, {Oliver},
  {Olivetto}, {Oppermann}, {Oram}, {O'Reilly}, {Ormiston}, {Ortega},
  {O'Shaughnessy}, {Ossokine}, {Osthelder}, {Ottaway}, {Overmier}, {Owen},
  {Pace}, {Pagano}, {Page}, {Pagliaroli}, {Pai}, {Pai}, {Palamos}, {Palashov},
  {Palomba}, {Pan}, {Panda}, {Pang}, {Pankow}, {Pannarale}, {Pant}, {Paoletti},
  {Paoli}, {Parida}, {Parker}, {Pascucci}, {Pasqualetti}, {Passaquieti},
  {Passuello}, {Patricelli}, {Payne}, {Pearlstone}, {Pechsiri}, {Pedersen},
  {Pedraza}, {Pele}, {Penn}, {Perego}, {Perez}, {P{\'e}rigois}, {Perreca},
  {Perri{\`e}s}, {Petermann}, {Pfeiffer}, {Phelps}, {Phukon}, {Piccinni},
  {Pichot}, {Piendibene}, {Piergiovanni}, {Pierro}, {Pillant}, {Pinard},
  {Pinto}, {Piotrzkowski}, {Pirello}, {Pitkin}, {Plastino}, {Poggiani}, {Pong},
  {Ponrathnam}, {Popolizio}, {Porter}, {Powell}, {Prajapati}, {Prasai},
  {Prasanna}, {Pratten}, {Prestegard}, {Principe}, {Prodi}, {Prokhorov},
  {Punturo}, {Puppo}, {P{\"u}rrer}, {Qi}, {Quetschke}, {Quinonez}, {Raab},
  {Raaijmakers}, {Radkins}, {Radulesco}, {Raffai}, {Rafferty}, {Raja}, {Rajan},
  {Rajbhandari}, {Rakhmanov}, {Ramirez}, {Ramos-Buades}, {Rana}, {Rao},
  {Rapagnani}, {Raymond}, {Razzano}, {Read}, {Regimbau}, {Rei}, {Reid},
  {Reitze}, {Rettegno}, {Ricci}, {Richardson}, {Richardson}, {Ricker},
  {Riemenschneider}, {Riles}, {Rizzo}, {Robertson}, {Robinet}, {Rocchi},
  {Rodriguez-Soto}, {Rolland}, {Rollins}, {Roma}, {Romanelli}, {Romano},
  {Romel}, {Romero-Shaw}, {Romie}, {Rose}, {Rose}, {Rose}, {Rosi{\'n}ska},
  {Rosofsky}, {Ross}, {Rowan}, {Rowlinson}, {Roy}, {Roy}, {Roy}, {Ruggi},
  {Rutins}, {Ryan}, {Sachdev}, {Sadecki}, {Sakellariadou}, {Salafia},
  {Salconi}, {Saleem}, {Samajdar}, {Sanchez}, {Sanchez}, {Sanchis-Gual},
  {Sanders}, {Santiago}, {Santos}, {Sarin}, {Sassolas}, {Sathyaprakash},
  {Sauter}, {Savage}, {Savant}, {Sawant}, {Sayah}, {Schaetzl}, {Schale},
  {Scheel}, {Scheuer}, {Schmidt}, {Schnabel}, {Schofield}, {Sch{\"o}nbeck},
  {Schreiber}, {Schulte}, {Schutz}, {Schwarm}, {Schwartz}, {Scott}, {Scott},
  {Seidel}, {Sellers}, {Sengupta}, {Sennett}, {Sentenac}, {Sequino}, {Sergeev},
  {Setyawati}, {Shaddock}, {Shaffer}, {Shahriar}, {Sharifi}, {Sharma},
  {Sharma}, {Shawhan}, {Shen}, {Shikauchi}, {Shink}, {Shoemaker}, {Shoemaker},
  {Shukla}, {ShyamSundar}, {Siellez}, {Sieniawska}, {Sigg}, {Singer}, {Singh},
  {Singh}, {Singha}, {Singhal}, {Sintes}, {Sipala}, {Skliris}, {Slagmolen},
  {Slaven-Blair}, {Smetana}, {Smith}, {Smith}, {Somala}, {Son}, {Soni},
  {Sorazu}, {Sordini}, {Sorrentino}, {Souradeep}, {Sowell}, {Spencer}, {Spera},
  {Srivastava}, {Srivastava}, {Staats}, {Stachie}, {Standke}, {Steer},
  {Steinke}, {Steinlechner}, {Steinlechner}, {Steinmeyer}, {Stevenson},
  {Stocks}, {Stops}, {Stover}, {Strain}, {Stratta}, {Strunk}, {Sturani},
  {Stuver}, {Sudhagar}, {Sudhir}, {Summerscales}, {Sun}, {Sunil}, {Sur},
  {Suresh}, {Sutton}, {Swinkels}, {Szczepa{\'n}czyk}, {Tacca}, {Tait},
  {Talbot}, {Tanasijczuk}, {Tanner}, {Tao}, {T{\'a}pai}, {Tapia}, {San Martin},
  {Tasson}, {Taylor}, {Tenorio}, {Terkowski}, {Thirugnanasambandam}, {Thomas},
  {Thomas}, {Thompson}, {Thondapu}, {Thorne}, {Thrane}, {Tinsman}, {Saravanan},
  {Tiwari}, {Tiwari}, {Tiwari}, {Toland}, {Tonelli}, {Tornasi},
  {Torres-Forn{\'e}}, {Torrie}, {Tosta e Melo}, {T{\"o}yr{\"a}}, {Trail},
  {Travasso}, {Traylor}, {Tringali}, {Tripathee}, {Trovato}, {Trudeau},
  {Tsang}, {Tse}, {Tso}, {Tsukada}, {Tsuna}, {Tsutsui}, {Turconi}, {Ubhi},
  {Udall}, {Ueno}, {Ugolini}, {Unnikrishnan}, {Urban}, {Usman}, {Utina},
  {Vahlbruch}, {Vajente}, {Valdes}, {Valentini}, {van Bakel}, {van Beuzekom},
  {van den Brand}, {Van Den Broeck}, {Vander-Hyde}, {van der Schaaf}, {Van
  Heijningen}, {van Veggel}, {Vardaro}, {Varma}, {Vass}, {Vas{\'u}th},
  {Vecchio}, {Vedovato}, {Veitch}, {Veitch}, {Venkateswara}, {Venugopalan},
  {Verkindt}, {Veske}, {Vetrano}, {Vicer{\'e}}, {Viets}, {Vinciguerra}, {Vine},
  {Vinet}, {Vitale}, {Vivanco}, {Vo}, {Vocca}, {Vorvick}, {Vyatchanin}, {Wade},
  {Wade}, {Wade}, {Walet}, {Walker}, {Wallace}, {Wallace}, {Walsh}, {Wang},
  {Wang}, {Wang}, {Ward}, {Warden}, {Warner}, {Was}, {Watchi}, {Weaver}, {Wei},
  {Weinert}, {Weinstein}, {Weiss}, {Wellmann}, {Wen}, {We{\ss}els},
  {Westhouse}, {Wette}, {Whelan}, {Whiting}, {Whittle}, {Wilken}, {Williams},
  {Willis}, {Willke}, {Winkler}, {Wipf}, {Wittel}, {Woan}, {Woehler},
  {Wofford}, {Wong}, {Wright}, {Wu}, {Wysocki}, {Xiao}, {Yamamoto}, {Yang},
  {Yang}, {Yang}, {Yap}, {Yazback}, {Yeeles}, {Yu}, {Yu}, {Yuen},
  {Zadro{\.z}ny}, {Zadro{\.z}ny}, {Zanolin}, {Zelenova}, {Zendri}, {Zevin},
  {Zhang}, {Zhang}, {Zhang}, {Zhao}, {Zhao}, {Zhou}, {Zhou}, {Zhu},
  {Zimmerman}, {Zlochower}, {Zucker}, {Zweizig}, {LIGO Scientific
  Collaboration}, \& {Virgo Collaboration}}]{Abbott2020ApJL}
---. 2020{\natexlab{b}}, \apjl, 900, L13, \dodoi{10.3847/2041-8213/aba493}

\bibitem[{{Acernese} {et~al.}(2015){Acernese}, {Agathos}, {Agatsuma}, {Aisa},
  {Allemandou}, {Allocca}, {Amarni}, {Astone}, {Balestri}, {Ballardin},
  {Barone}, {Baronick}, {Barsuglia}, {Basti}, {Basti}, {Bauer}, {Bavigadda},
  {Bejger}, {Beker}, {Belczynski}, {Bersanetti}, {Bertolini}, {Bitossi},
  {Bizouard}, {Bloemen}, {Blom}, {Boer}, {Bogaert}, {Bondi}, {Bondu},
  {Bonelli}, {Bonnand}, {Boschi}, {Bosi}, {Bouedo}, {Bradaschia}, {Branchesi},
  {Briant}, {Brillet}, {Brisson}, {Bulik}, {Bulten}, {Buskulic}, {Buy},
  {Cagnoli}, {Calloni}, {Campeggi}, {Canuel}, {Carbognani}, {Cavalier},
  {Cavalieri}, {Cella}, {Cesarini}, {Mottin}, {Chincarini}, {Chiummo}, {Chua},
  {Cleva}, {Coccia}, {Cohadon}, {Colla}, {Colombini}, {Conte}, {Coulon},
  {Cuoco}, {Dalmaz}, {D'Antonio}, {Dattilo}, {Davier}, {Day}, {Debreczeni},
  {Degallaix}, {Del{\'e}glise}, {Pozzo}, {Dereli}, {Rosa}, {Fiore}, {Lieto},
  {Virgilio}, {Doets}, {Dolique}, {Drago}, {Ducrot}, {Endr{\H{o}}czi},
  {Fafone}, {Farinon}, {Ferrante}, {Ferrini}, {Fidecaro}, {Fiori}, {Flaminio},
  {Fournier}, {Franco}, {Frasca}, {Frasconi}, {Gammaitoni}, {Garufi},
  {Gaspard}, {Gatto}, {Gemme}, {Gendre}, {Genin}, {Gennai}, {Ghosh},
  {Giacobone}, {Giazotto}, {Gouaty}, {Granata}, {Greco}, {Groot}, {Guidi},
  {Harms}, {Heidmann}, {Heitmann}, {Hello}, {Hemming}, {Hennes}, {Hofman},
  {Jaranowski}, {Jonker}, {Kasprzack}, {K{\'e}f{\'e}lian}, {Kowalska}, {Kraan},
  {Kr{\'o}lak}, {Kutynia}, {Lazzaro}, {Leonardi}, {Leroy}, {Letendre}, {Li},
  {Lieunard}, {Lorenzini}, {Loriette}, {Losurdo}, {Magazz{\`u}}, {Majorana},
  {Maksimovic}, {Malvezzi}, {Man}, {Mangano}, {Mantovani}, {Marchesoni},
  {Marion}, {Marque}, {Martelli}, {Martellini}, {Masserot}, {Meacher},
  {Meidam}, {Mezzani}, {Michel}, {Milano}, {Minenkov}, {Moggi}, {Mohan},
  {Montani}, {Morgado}, {Mours}, {Mul}, {Nagy}, {Nardecchia}, {Naticchioni},
  {Nelemans}, {Neri}, {Neri}, {Nocera}, {Pacaud}, {Palomba}, {Paoletti},
  {Paoli}, {Pasqualetti}, {Passaquieti}, {Passuello}, {Perciballi}, {Petit},
  {Pichot}, {Piergiovanni}, {Pillant}, {Piluso}, {Pinard}, {Poggiani},
  {Prijatelj}, {Prodi}, {Punturo}, {Puppo}, {Rabeling}, {R{\'a}cz},
  {Rapagnani}, {Razzano}, {Re}, {Regimbau}, {Ricci}, {Robinet}, {Rocchi},
  {Rolland}, {Romano}, {Rosi{\'n}ska}, {Ruggi}, {Saracco}, {Sassolas},
  {Schimmel}, {Sentenac}, {Sequino}, {Shah}, {Siellez}, {Straniero},
  {Swinkels}, {Tacca}, {Tonelli}, {Travasso}, {Turconi}, {Vajente}, {van
  Bakel}, {van Beuzekom}, {van den Brand}, {Van Den Broeck}, {van der Sluys},
  {van Heijningen}, {Vas{\'u}th}, {Vedovato}, {Veitch}, {Verkindt}, {Vetrano},
  {Vicer{\'e}}, {Vinet}, {Visser}, {Vocca}, {Ward}, {Was}, {Wei}, {Yvert},
  {{\.z}ny}, \& {Zendri}}]{Acernese2015}
{Acernese}, F., {Agathos}, M., {Agatsuma}, K., {et~al.} 2015, Classical and
  Quantum Gravity, 32, 024001, \dodoi{10.1088/0264-9381/32/2/024001}

\bibitem[{{Antonini} {et~al.}(2019){Antonini}, {Gieles}, \&
  {Gualandris}}]{Antonini2019MNRAS}
{Antonini}, F., {Gieles}, M., \& {Gualandris}, A. 2019, \mnras, 486, 5008,
  \dodoi{10.1093/mnras/stz1149}

\bibitem[{{Barber} \& {Antonini}(2024)}]{Barber2024arXiv}
{Barber}, J., \& {Antonini}, F. 2024, arXiv e-prints, arXiv:2410.03832,
  \dodoi{10.48550/arXiv.2410.03832}

\bibitem[{{Bartos} {et~al.}(2017){Bartos}, {Haiman}, {Marka}, {Metzger},
  {Stone}, \& {Marka}}]{Bartos2017NatCo}
{Bartos}, I., {Haiman}, Z., {Marka}, Z., {et~al.} 2017, Nature Communications,
  8, 831, \dodoi{10.1038/s41467-017-00851-7}

\bibitem[{{Baruteau} {et~al.}(2011){Baruteau}, {Cuadra}, \&
  {Lin}}]{Baruteau2011ApJ}
{Baruteau}, C., {Cuadra}, J., \& {Lin}, D.~N.~C. 2011, \apj, 726, 28,
  \dodoi{10.1088/0004-637X/726/1/28}

\bibitem[{{Belczynski} {et~al.}(2016){Belczynski}, {Holz}, {Bulik}, \&
  {O'Shaughnessy}}]{Belczynski2016Natur}
{Belczynski}, K., {Holz}, D.~E., {Bulik}, T., \& {O'Shaughnessy}, R. 2016,
  \nat, 534, 512, \dodoi{10.1038/nature18322}

\bibitem[{{Bellovary} {et~al.}(2016){Bellovary}, {Mac Low}, {McKernan}, \&
  {Ford}}]{Bellovary2016ApJ}
{Bellovary}, J.~M., {Mac Low}, M.-M., {McKernan}, B., \& {Ford}, K.~E.~S. 2016,
  \apjl, 819, L17, \dodoi{10.3847/2041-8205/819/2/L17}

\bibitem[{{Bethe} \& {Brown}(1998)}]{Bethe1998ApJ}
{Bethe}, H.~A., \& {Brown}, G.~E. 1998, \apj, 506, 780, \dodoi{10.1086/306265}

\bibitem[{{Calcino} {et~al.}(2024){Calcino}, {Dempsey}, {Dittmann}, \&
  {Li}}]{Calcino2024ApJ}
{Calcino}, J., {Dempsey}, A.~M., {Dittmann}, A.~J., \& {Li}, H. 2024, \apj,
  970, 107, \dodoi{10.3847/1538-4357/ad4a53}

\bibitem[{{Chen} {et~al.}(2022){Chen}, {Haster}, {Vitale}, {Farr}, \&
  {Isi}}]{Chen2022MNRAS}
{Chen}, H.-Y., {Haster}, C.-J., {Vitale}, S., {Farr}, W.~M., \& {Isi}, M. 2022,
  \mnras, 513, 2152, \dodoi{10.1093/mnras/stac989}

\bibitem[{{Dempsey} {et~al.}(2022){Dempsey}, {Li}, {Mishra}, \&
  {Li}}]{Dempsey2022ApJ}
{Dempsey}, A.~M., {Li}, H., {Mishra}, B., \& {Li}, S. 2022, \apj, 940, 155,
  \dodoi{10.3847/1538-4357/ac9d92}

\bibitem[{{Dodici} \& {Tremaine}(2024)}]{Dodici2024ApJ}
{Dodici}, M., \& {Tremaine}, S. 2024, \apj, 972, 193,
  \dodoi{10.3847/1538-4357/ad5cf2}

\bibitem[{{Everhart}(1985)}]{Everhart1985ASSL}
{Everhart}, E. 1985, in Astrophysics and Space Science Library, Vol. 115, IAU
  Colloq. 83: Dynamics of Comets: Their Origin and Evolution, ed. A.~{Carusi}
  \& G.~B. {Valsecchi}, 185, \dodoi{10.1007/978-94-009-5400-7_17}

\bibitem[{{Fabj} \& {Samsing}(2024)}]{Fabj2024arXiv}
{Fabj}, G., \& {Samsing}, J. 2024, arXiv e-prints, arXiv:2402.16948,
  \dodoi{10.48550/arXiv.2402.16948}

\bibitem[{{Gayathri} {et~al.}(2020){Gayathri}, {Bartos}, {Haiman}, {Klimenko},
  {Kocsis}, {M{\'a}rka}, \& {Yang}}]{Gayathri2020ApJ}
{Gayathri}, V., {Bartos}, I., {Haiman}, Z., {et~al.} 2020, \apjl, 890, L20,
  \dodoi{10.3847/2041-8213/ab745d}

\bibitem[{{Giacobbo} \& {Mapelli}(2018)}]{Giacobbo2018MNRAS}
{Giacobbo}, N., \& {Mapelli}, M. 2018, \mnras, 480, 2011,
  \dodoi{10.1093/mnras/sty1999}

\bibitem[{{Ginat} \& {Perets}(2021)}]{Giant2021MNRAS}
{Ginat}, Y.~B., \& {Perets}, H.~B. 2021, \mnras, 508, 190,
  \dodoi{10.1093/mnras/stab2565}

\bibitem[{{Ginat} \& {Perets}(2023)}]{Ginat2023MNRAS}
---. 2023, \mnras, 519, L15, \dodoi{10.1093/mnrasl/slac145}

\bibitem[{{Goldreich} \& {Tremaine}(1979)}]{Goldreich1979ApJ}
{Goldreich}, P., \& {Tremaine}, S. 1979, \apj, 233, 857, \dodoi{10.1086/157448}

\bibitem[{{Graham} {et~al.}(2020){Graham}, {Ford}, {McKernan}, {Ross}, {Stern},
  {Burdge}, {Coughlin}, {Djorgovski}, {Drake}, {Duev}, {Kasliwal}, {Mahabal},
  {van Velzen}, {Belecki}, {Bellm}, {Burruss}, {Cenko}, {Cunningham}, {Helou},
  {Kulkarni}, {Masci}, {Prince}, {Reiley}, {Rodriguez}, {Rusholme}, {Smith}, \&
  {Soumagnac}}]{Graham2020PhRvL}
{Graham}, M.~J., {Ford}, K.~E.~S., {McKernan}, B., {et~al.} 2020, \prl, 124,
  251102, \dodoi{10.1103/PhysRevLett.124.251102}

\bibitem[{{G{\"u}ltekin} {et~al.}(2006){G{\"u}ltekin}, {Miller}, \&
  {Hamilton}}]{Gultekin2006ApJ}
{G{\"u}ltekin}, K., {Miller}, M.~C., \& {Hamilton}, D.~P. 2006, \apj, 640, 156,
  \dodoi{10.1086/499917}

\bibitem[{{Hawley} {et~al.}(1995){Hawley}, {Gammie}, \&
  {Balbus}}]{Hawley1995ApJ}
{Hawley}, J.~F., {Gammie}, C.~F., \& {Balbus}, S.~A. 1995, \apj, 440, 742,
  \dodoi{10.1086/175311}

\bibitem[{{Kagra Collaboration} {et~al.}(2019){Kagra Collaboration}, {Akutsu},
  {Ando}, {Arai}, {Arai}, {Araki}, {Araya}, {Aritomi}, {Asada}, {Aso},
  {Atsuta}, {Awai}, {Bae}, {Baiotti}, {Barton}, {Cannon}, {Capocasa}, {Chen},
  {Chiu}, {Cho}, {Chu}, {Craig}, {Creus}, {Doi}, {Eda}, {Enomoto}, {Flaminio},
  {Fujii}, {Fujimoto}, {Fukunaga}, {Fukushima}, {Furuhata}, {Haino},
  {Hasegawa}, {Hashino}, {Hayama}, {Hirobayashi}, {Hirose}, {Hsieh}, {Huang},
  {Ikenoue}, {Inoue}, {Ioka}, {Itoh}, {Izumi}, {Kaji}, {Kajita}, {Kakizaki},
  {Kamiizumi}, {Kanbara}, {Kanda}, {Kanemura}, {Kaneyama}, {Kang}, {Kasuya},
  {Kataoka}, {Kawai}, {Kawamura}, {Kawasaki}, {Kim}, {Kim}, {Kim}, {Kim},
  {Kim}, {Kimura}, {Kinugawa}, {Kirii}, {Kitaoka}, {Kitazawa}, {Kojima},
  {Kokeyama}, {Komori}, {Kong}, {Kotake}, {Kozu}, {Kumar}, {Kuo}, {Kuroyanagi},
  {Lee}, {Lee}, {Lee}, {Leonardi}, {Lin}, {Lin}, {Liu}, {Liu}, {Majorana},
  {Mano}, {Marchio}, {Matsui}, {Matsushima}, {Michimura}, {Mio}, {Miyakawa},
  {Miyamoto}, {Miyamoto}, {Miyo}, {Miyoki}, {Morii}, {Morisaki}, {Moriwaki},
  {Morozumi}, {Musha}, {Nagano}, {Nagano}, {Nakamura}, {Nakamura}, {Nakano},
  {Nakano}, {Nakao}, {Narikawa}, {Naticchioni}, {Nguyen Quynh}, {Ni},
  {Nishizawa}, {Obuchi}, {Ochi}, {Oh}, {Oh}, {Ohashi}, {Ohishi}, {Ohkawa},
  {Okutomi}, {Ono}, {Oohara}, {Ooi}, {Pan}, {Park}, {Pe{\~n}a Arellano},
  {Pinto}, {Sago}, {Saijo}, {Saitou}, {Saito}, {Sakai}, {Sakai}, {Sakai},
  {Sasai}, {Sasaki}, {Sasaki}, {Sato}, {Sato}, {Sato}, {Sekiguchi}, {Seto},
  {Shibata}, {Shimoda}, {Shinkai}, {Shishido}, {Shoda}, {Somiya}, {Son},
  {Suemasa}, {Suzuki}, {Suzuki}, {Tagoshi}, {Tahara}, {Takahashi}, {Takahashi},
  {Takamori}, {Takeda}, {Tanaka}, {Tanaka}, {Tanaka}, {Tanioka}, {Tapia San
  Martin}, {Tatsumi}, {Tomaru}, {Tomura}, {Travasso}, {Tsubono}, {Tsuchida},
  {Uchikata}, {Uchiyama}, {Uehara}, {Ueki}, {Ueno}, {Uraguchi}, {Ushiba}, {van
  Putten}, {Vocca}, {Wada}, {Wakamatsu}, {Watanabe}, {Xu}, {Yamada},
  {Yamamoto}, {Yamamoto}, {Yamamoto}, {Yamamoto}, {Yamamoto}, {Yokogawa},
  {Yokoyama}, {Yokozawa}, {Yoon}, {Yoshioka}, {Yuzurihara}, {Zeidler}, \&
  {Zhu}}]{Kagra2019NatAs}
{Kagra Collaboration}, {Akutsu}, T., {Ando}, M., {et~al.} 2019, Nature
  Astronomy, 3, 35, \dodoi{10.1038/s41550-018-0658-y}

\bibitem[{{Kennedy} {et~al.}(2016){Kennedy}, {Meiron}, {Shukirgaliyev},
  {Panamarev}, {Berczik}, {Just}, \& {Spurzem}}]{Kennedy2016MNRAS}
{Kennedy}, G.~F., {Meiron}, Y., {Shukirgaliyev}, B., {et~al.} 2016, \mnras,
  460, 240, \dodoi{10.1093/mnras/stw908}

\bibitem[{{Kritos} \& {Silk}(2022)}]{Kritos2022PhRvD}
{Kritos}, K., \& {Silk}, J. 2022, \prd, 105, 063011,
  \dodoi{10.1103/PhysRevD.105.063011}

\bibitem[{{Li} \& {Fan}(2025)}]{Li-Fan2025arXiv}
{Li}, G.-P., \& {Fan}, X.-L. 2025, arXiv e-prints, arXiv:2502.11489.
\newblock \doarXiv{2502.11489}

\bibitem[{{Li} {et~al.}(2023){Li}, {Dempsey}, {Li}, {Lai}, \&
  {Li}}]{LJR2023ApJ}
{Li}, J., {Dempsey}, A.~M., {Li}, H., {Lai}, D., \& {Li}, S. 2023, \apjl, 944,
  L42, \dodoi{10.3847/2041-8213/acb934}

\bibitem[{{Li} \& {Lai}(2022)}]{LRX2022MNRAS}
{Li}, R., \& {Lai}, D. 2022, \mnras, 517, 1602, \dodoi{10.1093/mnras/stac2577}

\bibitem[{{Li} {et~al.}(2024){Li}, {Chen}, \& {Lin}}]{LYP2024ApJ}
{Li}, Y.-P., {Chen}, Y.-X., \& {Lin}, D. N.~C. 2024, \apj, 971, 130,
  \dodoi{10.3847/1538-4357/ad5a06}

\bibitem[{{Li} {et~al.}(2022){Li}, {Dempsey}, {Li}, {Li}, \& {Li}}]{LYP2022ApJ}
{Li}, Y.-P., {Dempsey}, A.~M., {Li}, H., {Li}, S., \& {Li}, J. 2022, \apjl,
  928, L19, \dodoi{10.3847/2041-8213/ac60fd}

\bibitem[{{Li} {et~al.}(2021){Li}, {Dempsey}, {Li}, {Li}, \& {Li}}]{LYP2021ApJ}
{Li}, Y.-P., {Dempsey}, A.~M., {Li}, S., {Li}, H., \& {Li}, J. 2021, \apj, 911,
  124, \dodoi{10.3847/1538-4357/abed48}

\bibitem[{{LIGO Scientific Collaboration} {et~al.}(2015){LIGO Scientific
  Collaboration}, {Aasi}, {Abbott}, {Abbott}, {Abbott}, {Abernathy}, {Ackley},
  {Adams}, {Adams}, {Addesso}, {Adhikari}, {Adya}, {Affeldt}, {Aggarwal},
  {Aguiar}, {Ain}, {Ajith}, {Alemic}, {Allen}, {Amariutei}, {Anderson},
  {Anderson}, {Arai}, {Araya}, {Arceneaux}, {Areeda}, {Ashton}, {Ast}, {Aston},
  {Aufmuth}, {Aulbert}, {Aylott}, {Babak}, {Baker}, {Ballmer}, {Barayoga},
  {Barbet}, {Barclay}, {Barish}, {Barker}, {Barr}, {Barsotti}, {Bartlett},
  {Barton}, {Bartos}, {Bassiri}, {Batch}, {Baune}, {Behnke}, {Bell}, {Bell},
  {Benacquista}, {Bergman}, {Bergmann}, {Berry}, {Betzwieser}, {Bhagwat},
  {Bhandare}, {Bilenko}, {Billingsley}, {Birch}, {Biscans}, {Biwer},
  {Blackburn}, {Blackburn}, {Blair}, {Blair}, {Bock}, {Bodiya}, {Bojtos},
  {Bond}, {Bork}, {Born}, {Bose}, {Brady}, {Braginsky}, {Brau}, {Bridges},
  {Brinkmann}, {Brooks}, {Brown}, {Brown}, {Brown}, {Buchman}, {Buikema},
  {Buonanno}, {Cadonati}, {Calder{\'o}n Bustillo}, {Camp}, {Cannon}, {Cao},
  {Capano}, {Caride}, {Caudill}, {Cavagli{\`a}}, {Cepeda}, {Chakraborty},
  {Chalermsongsak}, {Chamberlin}, {Chao}, {Charlton}, {Chen}, {Cho}, {Cho},
  {Chow}, {Christensen}, {Chu}, {Chung}, {Ciani}, {Clara}, {Clark}, {Collette},
  {Cominsky}, {Constancio}, {Cook}, {Corbitt}, {Cornish}, {Corsi}, {Costa},
  {Coughlin}, {Countryman}, {Couvares}, {Coward}, {Cowart}, {Coyne}, {Coyne},
  {Craig}, {Creighton}, {Creighton}, {Cripe}, {Crowder}, {Cumming},
  {Cunningham}, {Cutler}, {Dahl}, {Dal Canton}, {Damjanic}, {Danilishin},
  {Danzmann}, {Dartez}, {Dave}, {Daveloza}, {Davies}, {Daw}, {DeBra}, {Del
  Pozzo}, {Denker}, {Dent}, {Dergachev}, {DeRosa}, {DeSalvo}, {Dhurandhar},
  {D{\textasciiacute}{\i}az}, {Di Palma}, {Dojcinoski}, {Dominguez}, {Donovan},
  {Dooley}, {Doravari}, {Douglas}, {Downes}, {Driggers}, {Du}, {Dwyer},
  {Eberle}, {Edo}, {Edwards}, {Edwards}, {Effler}, {Eggenstein}, {Ehrens},
  {Eichholz}, {Eikenberry}, {Essick}, {Etzel}, {Evans}, {Evans},
  {Factourovich}, {Fairhurst}, {Fan}, {Fang}, {Farr}, {Farr}, {Favata}, {Fays},
  {Fehrmann}, {Fejer}, {Feldbaum}, {Ferreira}, {Fisher}, {Frei}, {Freise},
  {Frey}, {Fricke}, {Fritschel}, {Frolov}, {Fuentes-Tapia}, {Fulda}, {Fyffe},
  {Gair}, {Gaonkar}, {Gehrels}, {Gergely}, {Giaime}, {Giardina}, {Gleason},
  {Goetz}, {Goetz}, {Gondan}, {Gonz{\'a}lez}, {Gordon}, {Gorodetsky}, {Gossan},
  {Go{\ss}ler}, {Gr{\"a}f}, {Graff}, {Grant}, {Gras}, {Gray}, {Greenhalgh},
  {Gretarsson}, {Grote}, {Grunewald}, {Guido}, {Guo}, {Gushwa}, {Gustafson},
  {Gustafson}, {Hacker}, {Hall}, {Hammond}, {Hanke}, {Hanks}, {Hanna},
  {Hannam}, {Hanson}, {Hardwick}, {Harry}, {Harry}, {Hart}, {Hartman},
  {Haster}, {Haughian}, {Hee}, {Heintze}, {Heinzel}, {Hendry}, {Heng},
  {Heptonstall}, {Heurs}, {Hewitson}, {Hild}, {Hoak}, {Hodge}, {Hollitt},
  {Holt}, {Hopkins}, {Hosken}, {Hough}, {Houston}, {Howell}, {Hu}, {Huerta},
  {Hughey}, {Husa}, {Huttner}, {Huynh}, {Huynh-Dinh}, {Idrisy}, {Indik},
  {Ingram}, {Inta}, {Islas}, {Isler}, {Isogai}, {Iyer}, {Izumi}, {Jacobson},
  {Jang}, {Jawahar}, {Ji}, {Jim{\'e}nez-Forteza}, {Johnson}, {Jones}, {Jones},
  {Ju}, {Haris}, {Kalogera}, {Kandhasamy}, {Kang}, {Kanner}, {Katsavounidis},
  {Katzman}, {Kaufer}, {Kaufer}, {Kaur}, {Kawabe}, {Kawazoe}, {Keiser},
  {Keitel}, {Kelley}, {Kells}, {Keppel}, {Key}, {Khalaidovski}, {Khalili},
  {Khazanov}, {Kim}, {Kim}, {Kim}, {Kim}, {Kim}, {King}, {King}, {Kinzel},
  {Kissel}, {Klimenko}, {Kline}, {Koehlenbeck}, {Kokeyama}, {Kondrashov},
  {Korobko}, {Korth}, {Kozak}, {Kringel}, {Krishnan}, {Krueger}, {Kuehn},
  {Kumar}, {Kumar}, {Kuo}, {Landry}, {Lantz}, {Larson}, {Lasky}, {Lazzarini},
  {Lazzaro}, {Le}, {Leaci}, {Leavey}, {Lebigot}, {Lee}, {Lee}, {Lee}, {Leong},
  {Levin}, {Levine}, {Lewis}, {Li}, {Libbrecht}, {Libson}, {Lin}, {Littenberg},
  {Lockerbie}, {Lockett}, {Logue}, {Lombardi}, {Lormand}, {Lough}, {Lubinski},
  {L{\"u}ck}, {Lundgren}, {Lynch}, {Ma}, {Macarthur}, {MacDonald},
  {Machenschalk}, {MacInnis}, {Macleod}, {Maga{\~n}a-Sandoval}, {Magee},
  {Mageswaran}, {Maglione}, {Mailand}, {Mandel}, {Mandic}, {Mangano},
  {Mansell}, {M{\'a}rka}, {M{\'a}rka}, {Markosyan}, {Maros}, {Martin},
  {Martin}, {Martynov}, {Marx}, {Mason}, {Massinger}, {Matichard}, {Matone},
  {Mavalvala}, {Mazumder}, {Mazzolo}, {McCarthy}, {McClelland}, {McCormick},
  {McGuire}, {McIntyre}, {McIver}, {McLin}, {McWilliams}, {Meadors},
  {Meinders}, {Melatos}, {Mendell}, {Mercer}, {Meshkov}, {Messenger}, {Meyers},
  {Miao}, {Middleton}, {Mikhailov}, {Miller}, {Miller}, {Millhouse}, {Ming},
  {Mirshekari}, {Mishra}, {Mitra}, {Mitrofanov}, {Mitselmakher}, {Mittleman},
  {Moe}, {Mohanty}, {Mohapatra}, {Moore}, {Moraru}, {Moreno}, {Morriss},
  {Mossavi}, {Mow-Lowry}, {Mueller}, {Mueller}, {Mukherjee}, {Mullavey},
  {Munch}, {Murphy}, {Murray}, {Mytidis}, {Nash}, {Nayak}, {Necula}, {Nedkova},
  {Newton}, {Nguyen}, {Nielsen}, {Nissanke}, {Nitz}, {Nolting}, {Normandin},
  {Nuttall}, {Ochsner}, {O'Dell}, {Oelker}, {Ogin}, {Oh}, {Oh}, {Ohme},
  {Oppermann}, {Oram}, {O'Reilly}, {Ortega}, {O'Shaughnessy}, {Osthelder},
  {Ott}, {Ottaway}, {Ottens}, {Overmier}, {Owen}, {Padilla}, {Pai}, {Pai},
  {Palashov}, {Pal-Singh}, {Pan}, {Pankow}, {Pannarale}, {Pant}, {Papa},
  {Paris}, {Patrick}, {Pedraza}, {Pekowsky}, {Pele}, {Penn}, {Perreca},
  {Phelps}, {Pierro}, {Pinto}, {Pitkin}, {Poeld}, {Post}, {Poteomkin},
  {Powell}, {Prasad}, {Predoi}, {Premachandra}, {Prestegard}, {Price},
  {Principe}, {Privitera}, {Prix}, {Prokhorov}, {Puncken}, {P{\"u}rrer}, {Qin},
  {Quetschke}, {Quintero}, {Quiroga}, {Quitzow-James}, {Raab}, {Rabeling},
  {Radkins}, {Raffai}, {Raja}, {Rajalakshmi}, {Rakhmanov}, {Ramirez},
  {Raymond}, {Reed}, {Reid}, {Reitze}, {Reula}, {Riles}, {Robertson}, {Robie},
  {Rollins}, {Roma}, {Romano}, {Romanov}, {Romie}, {Rowan}, {R{\"u}diger},
  {Ryan}, {Sachdev}, {Sadecki}, {Sadeghian}, {Saleem}, {Salemi}, {Sammut},
  {Sandberg}, {Sanders}, {Sannibale}, {Santiago-Prieto}, {Sathyaprakash},
  {Saulson}, {Savage}, {Sawadsky}, {Scheuer}, {Schilling}, {Schmidt},
  {Schnabel}, {Schofield}, {Schreiber}, {Schuette}, {Schutz}, {Scott}, {Scott},
  {Sellers}, {Sengupta}, {Sergeev}, {Serna}, {Sevigny}, {Shaddock}, {Shahriar},
  {Shaltev}, {Shao}, {Shapiro}, {Shawhan}, {Shoemaker}, {Sidery}, {Siemens},
  {Sigg}, {Silva}, {Simakov}, {Singer}, {Singer}, {Singh}, {Sintes},
  {Slagmolen}, {Smith}, {Smith}, {Smith}, {Smith-Lefebvre}, {Son}, {Sorazu},
  {Souradeep}, {Staley}, {Stebbins}, {Steinke}, {Steinlechner}, {Steinlechner},
  {Steinmeyer}, {Stephens}, {Steplewski}, {Stevenson}, {Stone}, {Strain},
  {Strigin}, {Sturani}, {Stuver}, {Summerscales}, {Sutton}, {Szczepanczyk},
  {Szeifert}, {Talukder}, {Tanner}, {T{\'a}pai}, {Tarabrin}, {Taracchini},
  {Taylor}, {Tellez}, {Theeg}, {Thirugnanasambandam}, {Thomas}, {Thomas},
  {Thorne}, {Thorne}, {Thrane}, {Tiwari}, {Tomlinson}, {Torres}, {Torrie},
  {Traylor}, {Tse}, {Tshilumba}, {Ugolini}, {Unnikrishnan}, {Urban}, {Usman},
  {Vahlbruch}, {Vajente}, {Valdes}, {Vallisneri}, {van Veggel}, {Vass},
  {Vaulin}, {Vecchio}, {Veitch}, {Veitch}, {Venkateswara}, {Vincent-Finley},
  {Vitale}, {Vo}, {Vorvick}, {Vousden}, {Vyatchanin}, {Wade}, {Wade}, {Wade},
  {Walker}, {Wallace}, {Walsh}, {Wang}, {Wang}, {Wang}, {Ward}, {Warner},
  {Was}, {Weaver}, {Weinert}, {Weinstein}, {Weiss}, {Welborn}, {Wen},
  {Wessels}, {Westphal}, {Wette}, {Whelan}, {Whitcomb}, {White}, {Whiting},
  {Wilkinson}, {Williams}, {Williams}, {Williamson}, {Willis}, {Willke},
  {Wimmer}, {Winkler}, {Wipf}, {Wittel}, {Woan}, {Worden}, {Xie}, {Yablon},
  {Yakushin}, {Yam}, {Yamamoto}, {Yancey}, {Yang}, {Zanolin}, {Zhang}, {Zhang},
  {Zhang}, {Zhang}, {Zhao}, {Zhou}, {Zhu}, {Zucker}, {Zuraw}, \&
  {Zweizig}}]{LIGO2015}
{LIGO Scientific Collaboration}, {Aasi}, J., {Abbott}, B.~P., {et~al.} 2015,
  Classical and Quantum Gravity, 32, 074001,
  \dodoi{10.1088/0264-9381/32/7/074001}

\bibitem[{{Lyra} {et~al.}(2010){Lyra}, {Paardekooper}, \& {Mac
  Low}}]{Lyra2010ApJ}
{Lyra}, W., {Paardekooper}, S.-J., \& {Mac Low}, M.-M. 2010, \apjl, 715, L68,
  \dodoi{10.1088/2041-8205/715/2/L68}

\bibitem[{{Masset} \& {Papaloizou}(2003)}]{Masset2003ApJ}
{Masset}, F.~S., \& {Papaloizou}, J.~C.~B. 2003, \apj, 588, 494,
  \dodoi{10.1086/373892}

\bibitem[{{McKernan} {et~al.}(2014){McKernan}, {Ford}, {Kocsis}, {Lyra}, \&
  {Winter}}]{McKernan2014MNRAS}
{McKernan}, B., {Ford}, K.~E.~S., {Kocsis}, B., {Lyra}, W., \& {Winter}, L.~M.
  2014, \mnras, 441, 900, \dodoi{10.1093/mnras/stu553}

\bibitem[{{McKernan} {et~al.}(2012){McKernan}, {Ford}, {Lyra}, \&
  {Perets}}]{McKernan2012MNRAS}
{McKernan}, B., {Ford}, K.~E.~S., {Lyra}, W., \& {Perets}, H.~B. 2012, \mnras,
  425, 460, \dodoi{10.1111/j.1365-2966.2012.21486.x}

\bibitem[{{Monaghan}(1976)}]{Monaghan1976MNRAS}
{Monaghan}, J.~J. 1976, \mnras, 177, 583, \dodoi{10.1093/mnras/177.3.583}

\bibitem[{{Parischewsky} {et~al.}(2023){Parischewsky}, {Trani}, \&
  {Leigh}}]{Parischewsky2023ScPPC}
{Parischewsky}, H.~D., {Trani}, A.~A., \& {Leigh}, N. W.~C. 2023, SciPost
  Physics Core, 6, 016, \dodoi{10.21468/SciPostPhysCore.6.1.016}

\bibitem[{{Portegies Zwart} \& {McMillan}(2000)}]{Zwart2000ApJ}
{Portegies Zwart}, S.~F., \& {McMillan}, S. L.~W. 2000, \apjl, 528, L17,
  \dodoi{10.1086/312422}

\bibitem[{{Rein} \& {Liu}(2012)}]{Rein2012A&A}
{Rein}, H., \& {Liu}, S.~F. 2012, \aap, 537, A128,
  \dodoi{10.1051/0004-6361/201118085}

\bibitem[{{Rein} \& {Spiegel}(2015)}]{Rein2015MNRAS}
{Rein}, H., \& {Spiegel}, D.~S. 2015, \mnras, 446, 1424,
  \dodoi{10.1093/mnras/stu2164}

\bibitem[{{Rodriguez} {et~al.}(2016{\natexlab{a}}){Rodriguez}, {Chatterjee}, \&
  {Rasio}}]{Rodriguez2016PhRvD}
{Rodriguez}, C.~L., {Chatterjee}, S., \& {Rasio}, F.~A. 2016{\natexlab{a}},
  \prd, 93, 084029, \dodoi{10.1103/PhysRevD.93.084029}

\bibitem[{{Rodriguez} {et~al.}(2016{\natexlab{b}}){Rodriguez}, {Haster},
  {Chatterjee}, {Kalogera}, \& {Rasio}}]{Rodriguez2016ApJ}
{Rodriguez}, C.~L., {Haster}, C.-J., {Chatterjee}, S., {Kalogera}, V., \&
  {Rasio}, F.~A. 2016{\natexlab{b}}, \apjl, 824, L8,
  \dodoi{10.3847/2041-8205/824/1/L8}

\bibitem[{{Rowan} {et~al.}(2023){Rowan}, {Boekholt}, {Kocsis}, \&
  {Haiman}}]{Rowan2023MNRAS}
{Rowan}, C., {Boekholt}, T., {Kocsis}, B., \& {Haiman}, Z. 2023, \mnras, 524,
  2770, \dodoi{10.1093/mnras/stad1926}

\bibitem[{{Rowan} {et~al.}(2025){Rowan}, {Whitehead}, {Fabj}, {Saini},
  {Kocsis}, {Pessah}, \& {Samsing}}]{Rowan2025arXiv}
{Rowan}, C., {Whitehead}, H., {Fabj}, G., {et~al.} 2025, arXiv e-prints,
  arXiv:2501.09017.
\newblock \doarXiv{2501.09017}

\bibitem[{{Rowan} {et~al.}(2024){Rowan}, {Whitehead}, \&
  {Kocsis}}]{Rowan2024arXiv}
{Rowan}, C., {Whitehead}, H., \& {Kocsis}, B. 2024, arXiv e-prints,
  arXiv:2412.12086, \dodoi{10.48550/arXiv.2412.12086}

\bibitem[{{Rozner} {et~al.}(2023){Rozner}, {Generozov}, \&
  {Perets}}]{Rozner2023MNRAS}
{Rozner}, M., {Generozov}, A., \& {Perets}, H.~B. 2023, \mnras, 521, 866,
  \dodoi{10.1093/mnras/stad603}

\bibitem[{{Rozner} \& {Perets}(2022)}]{Rozner2022ApJ}
{Rozner}, M., \& {Perets}, H.~B. 2022, \apj, 931, 149,
  \dodoi{10.3847/1538-4357/ac6d55}

\bibitem[{{Saini}(2024)}]{Saini2024MNRAS}
{Saini}, P. 2024, \mnras, 528, 833, \dodoi{10.1093/mnras/stae037}

\bibitem[{{Samsing}(2018)}]{Samsing2018PhRvD}
{Samsing}, J. 2018, \prd, 97, 103014, \dodoi{10.1103/PhysRevD.97.103014}

\bibitem[{{Samsing} {et~al.}(2022){Samsing}, {Bartos}, {D'Orazio}, {Haiman},
  {Kocsis}, {Leigh}, {Liu}, {Pessah}, \& {Tagawa}}]{Samsing2022Natur}
{Samsing}, J., {Bartos}, I., {D'Orazio}, D.~J., {et~al.} 2022, \nat, 603, 237,
  \dodoi{10.1038/s41586-021-04333-1}

\bibitem[{{Secunda} {et~al.}(2019){Secunda}, {Bellovary}, {Mac Low}, {Ford},
  {McKernan}, {Leigh}, {Lyra}, \& {S{\'a}ndor}}]{Secunda2019ApJ}
{Secunda}, A., {Bellovary}, J., {Mac Low}, M.-M., {et~al.} 2019, \apj, 878, 85,
  \dodoi{10.3847/1538-4357/ab20ca}

\bibitem[{{Sirko} \& {Goodman}(2003)}]{SG2003MNRAS}
{Sirko}, E., \& {Goodman}, J. 2003, \mnras, 341, 501,
  \dodoi{10.1046/j.1365-8711.2003.06431.x}

\bibitem[{{Stevenson} {et~al.}(2017){Stevenson}, {Vigna-G{\'o}mez}, {Mandel},
  {Barrett}, {Neijssel}, {Perkins}, \& {de Mink}}]{Stevenson2017NatCo}
{Stevenson}, S., {Vigna-G{\'o}mez}, A., {Mandel}, I., {et~al.} 2017, Nature
  Communications, 8, 14906, \dodoi{10.1038/ncomms14906}

\bibitem[{{Stone} {et~al.}(1996){Stone}, {Hawley}, {Gammie}, \&
  {Balbus}}]{Stone1996ApJ}
{Stone}, J.~M., {Hawley}, J.~F., {Gammie}, C.~F., \& {Balbus}, S.~A. 1996,
  \apj, 463, 656, \dodoi{10.1086/177280}

\bibitem[{{Stone} {et~al.}(2020){Stone}, {Tomida}, {White}, \&
  {Felker}}]{stone2020ApJS}
{Stone}, J.~M., {Tomida}, K., {White}, C.~J., \& {Felker}, K.~G. 2020, \apjs,
  249, 4, \dodoi{10.3847/1538-4365/ab929b}

\bibitem[{{Stone} \& {Leigh}(2019)}]{Stone2019Natur}
{Stone}, N.~C., \& {Leigh}, N. W.~C. 2019, \nat, 576, 406,
  \dodoi{10.1038/s41586-019-1833-8}

\bibitem[{{Stone} {et~al.}(2017){Stone}, {Metzger}, \&
  {Haiman}}]{stone2017MNRAS}
{Stone}, N.~C., {Metzger}, B.~D., \& {Haiman}, Z. 2017, \mnras, 464, 946,
  \dodoi{10.1093/mnras/stw2260}

\bibitem[{{Tagawa} {et~al.}(2020){Tagawa}, {Haiman}, \&
  {Kocsis}}]{Tagawa2020ApJ}
{Tagawa}, H., {Haiman}, Z., \& {Kocsis}, B. 2020, \apj, 898, 25,
  \dodoi{10.3847/1538-4357/ab9b8c}

\bibitem[{{Tanaka} {et~al.}(2002){Tanaka}, {Takeuchi}, \&
  {Ward}}]{Tanaka2002ApJ}
{Tanaka}, H., {Takeuchi}, T., \& {Ward}, W.~R. 2002, \apj, 565, 1257,
  \dodoi{10.1086/324713}

\bibitem[{{Thompson} {et~al.}(2005){Thompson}, {Quataert}, \&
  {Murray}}]{TQM2005ApJ}
{Thompson}, T.~A., {Quataert}, E., \& {Murray}, N. 2005, \apj, 630, 167,
  \dodoi{10.1086/431923}

\bibitem[{{Trani} {et~al.}(2024){Trani}, {Quaini}, \& {Colpi}}]{Trani2024A&A}
{Trani}, A.~A., {Quaini}, S., \& {Colpi}, M. 2024, \aap, 683, A135,
  \dodoi{10.1051/0004-6361/202347920}

\bibitem[{{Valtonen} \& {Karttunen}(2006)}]{Valtonen2006tbp}
{Valtonen}, M., \& {Karttunen}, H. 2006, {The Three-Body Problem}

\bibitem[{{Vilkoviskij} \& {Czerny}(2002)}]{Vilkoviskij2002A&A}
{Vilkoviskij}, E.~Y., \& {Czerny}, B. 2002, \aap, 387, 804,
  \dodoi{10.1051/0004-6361:20020255}

\bibitem[{{Wang} {et~al.}(2023){Wang}, {Ma}, \& {Wu}}]{Wang2023MNRAS}
{Wang}, M., {Ma}, Y., \& {Wu}, Q. 2023, \mnras, 520, 4502,
  \dodoi{10.1093/mnras/stad422}

\bibitem[{{Whitehead} {et~al.}(2024{\natexlab{a}}){Whitehead}, {Rowan},
  {Boekholt}, \& {Kocsis}}]{Whitehead2024MNRAS}
{Whitehead}, H., {Rowan}, C., {Boekholt}, T., \& {Kocsis}, B.
  2024{\natexlab{a}}, \mnras, 531, 4656, \dodoi{10.1093/mnras/stae1430}

\bibitem[{{Whitehead} {et~al.}(2024{\natexlab{b}}){Whitehead}, {Rowan},
  {Boekholt}, \& {Kocsis}}]{Whitehead2024MNRAS_disk_nova}
---. 2024{\natexlab{b}}, \mnras, 533, 1766, \dodoi{10.1093/mnras/stae1866}

\bibitem[{{Yang} {et~al.}(2019){Yang}, {Bartos}, {Gayathri}, {Ford}, {Haiman},
  {Klimenko}, {Kocsis}, {M{\'a}rka}, {M{\'a}rka}, {McKernan}, \&
  {O'Shaughnessy}}]{Yang2019PhRvL}
{Yang}, Y., {Bartos}, I., {Gayathri}, V., {et~al.} 2019, \prl, 123, 181101,
  \dodoi{10.1103/PhysRevLett.123.181101}

\bibitem[{{Zackay} {et~al.}(2021){Zackay}, {Dai}, {Venumadhav}, {Roulet}, \&
  {Zaldarriaga}}]{Zackay2021PhRvD}
{Zackay}, B., {Dai}, L., {Venumadhav}, T., {Roulet}, J., \& {Zaldarriaga}, M.
  2021, \prd, 104, 063030, \dodoi{10.1103/PhysRevD.104.063030}

\bibitem[{{Ziosi} {et~al.}(2014){Ziosi}, {Mapelli}, {Branchesi}, \&
  {Tormen}}]{Ziosi2014MNRAS}
{Ziosi}, B.~M., {Mapelli}, M., {Branchesi}, M., \& {Tormen}, G. 2014, \mnras,
  441, 3703, \dodoi{10.1093/mnras/stu824}

\end{thebibliography}
\bibliographystyle{aasjournal}
\end{document}